\documentclass[aps, prd, superscriptaddress, nofootinbib, preprintnumbers]{revtex4}
\usepackage{amsmath}
\usepackage{graphicx}

\newcommand{\chushi}[1]{}

\begin{document}
 \preprint{MISC-2012-11}
 \title{{\bf Walking techni-pions at LHC}
 \vspace{5mm}}
  \author{{Junji Jia}} \thanks{
      {\tt junjijia@kmi.nagoya-u.ac.jp}}
      \affiliation{ Kobayashi-Maskawa Institute for the Origin of Particles and 
the Universe (KMI) \\ 
 Nagoya University, Nagoya 464-8602, Japan.}
\author{Shinya Matsuzaki}\thanks{
      {\tt synya@cc.kyoto-su.ac.jp} }
      \affiliation{ Maskawa Institute for Science and Culture, Kyoto Sangyo University, Motoyama, Kamigamo, Kita-Ku, Kyoto 603-8555, Japan.}
\author{{Koichi Yamawaki}} \thanks{
      {\tt yamawaki@kmi.nagoya-u.ac.jp}}
      \affiliation{ Kobayashi-Maskawa Institute for the Origin of Particles and 
the Universe (KMI) \\ 
 Nagoya University, Nagoya 464-8602, Japan.}
\date{\today}

\begin{abstract}
We calculate techni-pion masses 
of the walking technicolor (WTC),  
by explicitly evaluating nontrivial contributions 
from various possible chiral breaking sources in a concrete WTC setting of the one-family model. 
Our explicit computation of the mass and the coupling in this concrete model setting reveals that 
the techni-pions are on the order of several hundred GeV in the region to be discovered at LHC.  
\end{abstract}
\maketitle

\section{Introduction}

Technicolor (TC)~\cite{Weinberg:1975gm,Farhi:1980xs,Yamawaki:1996vr} 
provides the dynamical origin of the electroweak (EW)  symmetry breaking 
by triggering  condensation of techni-fermion bilinear, without introduction of 
a fundamental Higgs boson as in the standard model (SM). 
However, the original version of TC~\cite{Weinberg:1975gm},  a naive scale-up version of QCD,  
has already been excluded due to the excessive flavor changing neutral currents (FCNC).

The solution to the FCNC problem 
was  given by the walking TC 
having large anomalous dimension $\gamma_m=1$ due to the 
scale-invariant (conformal) gauge dynamics with non-running coupling 
~\cite{Yamawaki:1985zg}: The coupling is actually slowly running (walking)  in a non-perturbative 
sense a la Miransky~\cite{Miransky:1984ef}. (Subsequently similar ideas were proposed without notion of the anomalous dimension and the scale
invariance~\cite{Akiba:1985rr} ). ~\footnote{
Another problem of the TC as a QCD scale-up is the electroweak constraints, so-called $S$ and $T$  parameters. 
This may also be improved in the  
walking TC~\cite{Appelquist:1991is,Harada:2005ru}.  
Even if the 
walking TC in isolation cannot overcome this problem, there still exist a possibility that the problem may be 
resolved in the combined dynamical system including the SM fermion mass generation such as the extended TC 
(ETC) dynamics~\cite{Dimopoulos:1979es}, 
in much the same way as the solution (``ideal fermion delocalization'')~\cite{Cacciapaglia:2004rb} in the Higgsless models which simultaneously
adjust $S$ and $T$ parameters by incorporating the SM fermion mass profile. 
}   
The mass of techni-fermion $m_F (={\cal O}(1{\rm TeV}))$ is generated dynamically in such a way that  $m_F$ near the critical coupling  $\alpha\simeq \alpha_c$  can be exponentially smaller than the cutoff $\Lambda$ (to be identified with the scale of ETC~\cite{Dimopoulos:1979es}  $\Lambda=\Lambda_{\rm ETC}={\cal O}(10^3-10^4{\rm TeV})$), so-called Miransky scaling~\cite{Miransky:1984ef}, closely tied with the conformal phase transition~\cite{Miransky:1996pd}. Then the walking behavior extends in a wide region  $m_F < p < \Lambda$.

Such a scale-invariant dynamics may be realized in a model  
having a large number of techni-fermion flavors ($N_{\rm TF}$), as exemplified by the large $N_F$ QCD which has an approximate infrared fixed point 
(IRFP)~\cite{Caswell:1974gg} in the two-loop beta function~\cite{Lane:1991qh,Appelquist:1996dq, Miransky:1996pd}: Thanks to the IRFP the two-loop coupling is 
almost non-running up till the scale $\Lambda_{\rm TC}$, an analogue of $\Lambda_{\rm QCD}$,  above which the coupling runs as in a usual asymptotically free gauge theory like QCD and hence $\Lambda_{\rm TC}$ plays the role of the cutoff. 
In the same a way as the scale-invariant case, $m_F$ can be generated much smaller than $\Lambda_{\rm TC}$ a la Miransky scaling and hence 
 the non-perturbative walking regime of the coupling 
extends in a wide region of 
energy scale, $m_F < p < \Lambda_{\rm TC}$, 
whereas the asymptotically free region $p>\Lambda_{\rm TC}$ will be 
embedded in an ETC~\cite{Dimopoulos:1979es}. 
The chiral condensate is enhanced 
by the large anomalous dimension of techni-fermion bilinear operator $\gamma_m \simeq1$, 
so that realistic masses of SM light fermions can be realized without suffering from the FCNC problem~\footnote{
The top mass is quite hard to be reproduced by the walking TC with anomalous dimension $\gamma_m \simeq 1$. 
Other dynamics such as the top quark condensate~\cite{Miransky:1988xi} may be required. 
However, it was found~\cite{Miransky:1988gk} that  if we include additional four-fermion interactions like strong ETC, 
the anomalous dimension becomes much larger $1<\gamma_m<2$,  
which can boost the ETC-origin mass to arbitrarily large up till  the techni-fermion mass scale (``strong ETC model'').  
(Subsequently the same effects  were also noted  without concept of the anomalous dimension~\cite{Matumoto:1989hf}.) }.

As a concrete realization of the walking TC which generally needs a large number of technifermion flavors we may consider the
Farhi-Susskind one-family model~\cite{Farhi:1980xs}, 
which consists of one-family of the techni-fermions (techni-quarks and -leptons) 
having the same SM gauge charges as those of the SM fermions (ordinary quarks and leptons). 
The global chiral symmetry breaking then gets enhanced from the minimal structure, 
$SU(2)_L \times SU(2)_R \to SU(2)_V$ as in the SM, to an extended one, 
$SU(8)_L \times SU(8)_R \to SU(8)_V$, 
and hence gives rise to the associated sixty pseudo-Nambu Goldstone bosons (``techni-pions"). 
(Three of the total sixty-three are eaten by the SM weak bosons.)  
Probing those techni-pions at collider experiments 
is thus necessary for discovering the walking TC.

The presence of the wide walking region implies 
approximate scale invariance, which is broken spontaneously by the techni-fermion mass generation at the same time 
the chiral symmetry is broken. 
The associated pseudo-Nambu Goldstone boson, ``techni-dilaton"~\cite{Yamawaki:1985zg}, 
therefore emerges as a light composite scalar formed as techni-fermion and anti-techni-fermion bound state. 
The walking low-lying spectra would thus consist of techni-pions and techni-dilaton (``walking pseudo's"), 
whose collider signatures would therefore serve as definite benchmarks toward discovery of the walking TC.

Actually, the techni-dilaton signatures at the LHC have recently been discussed~\cite{Matsuzaki:2011ie,Matsuzaki:2012gd,Matsuzaki:2012fq} 
in comparison with the SM Higgs. 
It was shown that the characteristic signatures are seen through 
the diphoton channel either at around 125 GeV consistently with the currently reported diphoton excess~\cite{ATLAS:2012ae} 
or above 600 GeV as a nonresonant excess in a higher energy region of the diphoton invariant mass distribution.

In Ref.~\cite{Chivukula:2011ue}, on the other hand, 
the current LHC limits on techni-pions have also been discussed focusing on 
isospin-color singlet techni-pion (denoted as $P^0$ in the original literature~\cite{Farhi:1980xs}) 
in the diphoton and tau lepton pair channels, taking the pion mass as a free parameter. 
In the case of the one-family walking TC, however, 
the techni-pion masses can be pulled up to a higher scale than that expected from naive 
scaling of QCD~\cite{Yamawaki:1996vr}. 
Such a naively believed folklore would be clarified by  
explicit estimate of techni-pion masses incorporating 
the essential features of walking dynamics, which would also 
find out more relevant parameter regions to search for walking 
techni-pions at the LHC. 
Actually, such explicit calculations have not been done so far.

In this paper, we compute the mass and the coupling of techni-pions 
of a typical walking TC, the Farhi-Susskind one-family model~\cite{Farhi:1980xs},
using recent results on a nonperturbative analysis based on 
the ladder Schwinger-Dyson equation employed in a modern version of walking TC~\cite{Hashimoto:2010nw}.  
The masses of techni-pions charged under the EW gauges 
are calculated by evaluating one-EW gauge boson exchange diagrams, so that 
the contributions are cast into the form of integral over the momentum square $Q^2$ 
with respect to difference between vector and axialvector current correlators $(\Pi_{V-A})$, similarly to 
computation for charged pion mass in QCD. 
Though those current correlators are quite sensitive to ultraviolet behavior and thus the walking dynamics, 
it turns out that the EW gauge boson exchange contributions dramatically cancel 
each other in the ultraviolet region, as was discussed long ago for the naive-scale up version of 
QCD~\cite{Eichten:1979ah,Peskin:1980gc,Chadha:1981yt}, so that there arise 
no sizable corrections to the masses.

The colored techni-pions, on the other hand,  get sizable ultraviolet contributions 
from one-gluon exchange diagram, in contrary to the charged pions. 
The size of corrections without ultraviolet cancellation is actually enhanced by a large logarithmic factor 
scaled with the ultraviolet scale of TC, $\Lambda_{\rm TC}$, 
compared to the naive-scale up version of QCD~\cite{Farhi:1980xs}. 
This is due to the characteristic ultraviolet scaling of $\Pi_{V-A}$ 
in the walking TC: the $\Pi_{V-A}$ in the walking TC  
damps with the large anomalous dimension $\gamma_m \simeq 1$ 
 more slowly than that in QCD-like dynamics with $\gamma_m \simeq 0$, 
in such a way that $\Pi_{V-A} \sim 1/Q^{4-2\gamma_m}$. 
Thus the amount of integration over the momentum square $Q^2$ 
gets larger than that in the case of QCD-like dynamics, depending on the size of the ultraviolet cutoff $\Lambda_{\rm TC}$,  
as was indicated in Ref.~\cite{Harada:2005ru}.

As in Ref.~\cite{Farhi:1980xs}, 
ETC-induced four-fermion interactions breaking the full chiral $SU(8)_L \times SU(8)_R$ symmetry into  
the separate chiral symmetries for techni-quarks and -leptons give the masses to techni-pions 
coupled to the separate chiral currents.  
The masses also get enhanced due to the chiral condensate enhanced by 
the large anomalous dimension, as has been expected~\cite{Yamawaki:1996vr}.  
Precise estimates of the masses can then be made by using the recent nonperturbative results~\cite{Hashimoto:2010nw}  
on the techni-fermion chiral condensate $\langle \bar{F}F  \rangle$ combined with the Pagels-Stokar formula~\cite{Pagels:1979hd} for the 
techni-pion decay constant $F_\pi$, which allows us to evaluate $\langle \bar{F}F  \rangle$ in terms of 
$F_\pi$ fixed as $F_\pi = v_{\rm EW}/2 \simeq 123$ GeV. 
As a result, it turns out that 
all the techni-pions are on the order of several hundred GeV (See Table~\ref{tab:TP}).

Based on our estimation, 
we then discuss the phenomenological implications to the LHC signatures, 
focusing on neutral isosinglet techni-pions, in comparison with the SM Higgs.

This paper is organized as follows: 
In Sec.~\ref{TPcouplings-masses} we start with a brief review of 
a low energy effective Lagrangian, which consists of the walking pseudo's 
(techni-pions and techni-dilaton), based on nonlinear 
realization of both chiral and scale symmetries~\cite{Matsuzaki:2011ie,Matsuzaki:2012gd,Matsuzaki:2012fq}.   
We then explicitly identify the techni-pion currents coupled to the SM gauge bosons and fermions 
and couplings necessary to 
calculate the techni-pion masses. 
The walking techni-pion masses are computed based on the standard current algebra. 
In Sec.~\ref{LHC} we address the phenomenological implications to the LHC 
focusing on neutral isosinglet techni-pions. 
Sec.~\ref{summary} is devoted to summary of this paper. 
In Appendix~\ref{sec:TD} we present a brief discussion about effects on techni-dilaton phenomenologies 
arising from the couplings to the techni-pions.

\section{The one-family walking techni-pion masses and couplings}
\label{TPcouplings-masses}

We begin with an effective Lagrangian 
relevant to the walking pseudo's (techni-pions and techni-dilaton)  
based on the nonlinear realization for both scale and chiral $SU(N_{\rm TF})_L \times SU(N_{\rm TF})_R$ 
symmetries~\cite{Matsuzaki:2011ie,Matsuzaki:2012gd,Matsuzaki:2012fq}. 
The building blocks consist of the usual chiral nonlinear base $U$ and techni-dilaton field $\phi$. 
The $U$ is parametrized as $U=e^{2i \pi/F_\pi}$ where $\pi=\pi^A X^A$ $(A=1,\cdots ,N_{\rm TF}^2-1)$ with $X^A$ being 
generators of $SU(N_{\rm TF})$ and $F_\pi$ denotes the decay constant associated with the spontaneous 
breaking of the chiral symmetry. 
The $U$ then transforms under the chiral symmetry as $U \to g_L U g_R^\dag$ where 
$g_{L,R} \in SU(N_{\rm TF})_{L,R}$, 
while under the scale symmetry $\delta U= x^\nu \partial_\nu U$ so does $\pi$. 
The techni-dilaton field $\phi$ is, on the other hand, introduced so as to parametrize 
a nonlinear base for the scale symmetry, $\chi$, such that $\chi=e^{\phi/F_{\phi}}$ 
with the decay constant for the spontaneous breaking of the scale symmetry $F_{\phi}$. 
The scale nonlinear base $\chi$ then transforms with scale dimension 1, i.e., $\delta \chi = (1 + x^\nu \partial_\nu) \chi$ 
so that $\phi$ does nonlinearly as $\delta \phi= F_{\phi} + x^\nu \partial_\nu \phi$.

One thus constructs the nonlinear Lagrangian~\cite{Matsuzaki:2011ie,Matsuzaki:2012gd,Matsuzaki:2012fq}:  
\begin{eqnarray} 
{\cal L}
&=& \frac{F_{\pi}^2}{4} \chi^2 {\rm tr}[{\cal D}_\mu U^\dag {\cal D}^\mu U] 
+ {\cal L}_{\pi ff} + 
\cdots \label{Lag:int} 
\end{eqnarray}
where 
${\cal D}_\mu U$ denotes the covariant derivative acting on $U$ gauged only under the SM $SU(3)_c \times SU(2)_W \times U(1)_Y$ 
gauge symmetries, which will later be specified to the case of the one-family model. 
The Yukawa interaction terms between the techni-pions and SM fermions are included in ${\cal L}_{\pi ff}$ 
which should involve a ``spurion field" $S(x)$~\cite{Matsuzaki:2011ie,Matsuzaki:2012gd,Matsuzaki:2012fq} necessary to 
reflect the explicit breaking of the scale symmetry due to the dynamical mass generation of techni-fermion.   
Actually, the Yukawa couplings highly depend on modeling of ETC, 
which arise necessary through the ETC-gauge boson exchanges.
We will later discuss the Yukawa couplings to fix the form  
by considering typical ETC exchange contributions. 
The ellipses in Eq.(\ref{Lag:int}) include techni-pion mass terms which are to be studied later.

\subsection{Techni-pion couplings}

In the Farhi-Susskind one-family model~\cite{Farhi:1980xs}, 
the chiral symmetry gets enhanced from the minimal $SU(2)_L \times SU(2)_R$ to 
$SU(2 N_D)_L \times SU(2 N_D)_R$, where $N_D=4$ corresponding to 
three techni-quark $Q_c$ ($c=r,g,b$) and one techni-lepton ($L$) doublets. 
The techni-fermion condensation $\langle \bar{F}F \rangle \neq 0$ ($F=Q, L$) therefore 
breaks the enlarged chiral symmetry down to $SU(8)_V$, leading to sixty-three Nambu-Goldstone bosons in total. 
The three of them become unphysical to be eaten by $W$ and $Z$ bosons in the same way as in 
the usual Higgs mechanism, while the other sixty Nambu-Goldstone bosons become pseudos, techni-pions, to be massive in several ways. 
The techni-pions are classified  by the isospin and QCD color charges, 
which are listed in Table~\ref{tab:TP} together with the characterized currents coupled to them, 
where the notation follows the original literature~\cite{Farhi:1980xs}.

\begin{table}
\begin{tabular}{c|c|c}
\hline 
\hspace{15pt} techni-pion \hspace{15pt} & \hspace{15pt} current \hspace{15pt} 
& \hspace{10pt} mass [GeV] (walking TC) \hspace{10pt} \\  
\hline 
$\theta_a^{i}$ & $\frac{1}{\sqrt{2}}\bar{Q} \gamma_\mu \gamma_5 \lambda_a \tau^{i} Q$ & $449 (537) \sqrt{ \frac{3}{N_{\rm TC}}} $ \\ 
$\theta_a$ &  $ \frac{1}{2\sqrt{2}} \bar{Q} \gamma_\mu \gamma_5 \lambda_a Q$ &   $449 (537) \sqrt{ \frac{3}{N_{\rm TC}}} $ \\ 
$T_c^{i}$ $(\bar{T}_c^{i})$ & $\frac{1}{\sqrt{2}} \bar{Q}_c \gamma_\mu \gamma_5 \tau^{i} L$ (h.c.) & $299 (358) \sqrt{ \frac{3}{N_{\rm TC}}} $  \\ 
$T_c$ ($\bar{T}_c$) & $ \frac{1}{2 \sqrt{2}} \bar{Q}_c \gamma_\mu \gamma_5 L $ (h.c.)   & $299 (358) \sqrt{ \frac{3}{N_{\rm TC}}} $ \\ 
\hline 
$P^i $ & $\frac{1}{2 \sqrt{3}} (\bar{Q} \gamma_\mu \gamma_5 \tau^i Q - 3 \bar{L} \gamma_\mu \gamma_5 \tau^i L)$ & 502 (ETC) \\ 
$P^0$ &  $\frac{1}{4 \sqrt{3}} (\bar{Q} \gamma_\mu \gamma_5  Q - 3 \bar{L} \gamma_\mu \gamma_5  L)$  & 397 (ETC) \\ 
\hline 
\end{tabular} 
\caption{ The techni-pions, their associated currents and masses in the original one-family model~\cite{Farhi:1980xs}. 
The masses have been estimated including the enhancement of techni-fermion condensation in the case of 
walking TC with $\Lambda_{\rm TC}=10^3\, (10^4)$ TeV (See text). 
Here $\lambda_a$ ($a=1,\cdots,8$) are the Gell-Mann matrices, 
$\tau^i$ $SU(2)$ generators normalized as 
$\tau^i=\sigma^i/2$ ($i=1,2,3$) with the Pauli matrices $\sigma^i$, and 
the label $c$ attached on the color-triplet techni-pion field $T_c$ stands for QCD-three colors, $c=r,g,b$. 
}
\label{tab:TP}
\end{table}

The techni-pion couplings in the one-family model are read off from Eq.(\ref{Lag:int}) 
once the broken generators $X^A$ ($A=1, \cdots , 63$) appropriate to the corresponding broken currents listed in Table~\ref{tab:TP}
are specified: 
\begin{eqnarray} 
\sum_{A=1}^{63} \pi^A(x) X^A 
&=& 
\sum_{i=1}^3 \pi_{\rm eaten}^i(x) X_{\rm eaten}^i  
+ 
\sum_{i=1}^3 P^i(x) X^i_{P} + P^0 (x) X_{P} 
\nonumber \\  
&& 
+ \sum_{i=1}^3 \sum_{a=1}^8 \theta^i_a (x) X_{\theta a}^i 
+ 
\sum_{a=1}^8 \theta_a (x) X_{\theta a}  
\nonumber \\ 
&& 
+ 
\sum_{c=r,g,b} \sum_{i=1}^3 \left[ T_c^{(1)i}(x) X_{T c}^{(1)i} + T_c^{(2)i}(x) X_{T c}^{(2)i} \right] 
+
\sum_{c=r,g,b} \left[ T_c^{(1)} (x) X_{T c}^{(1)}  + T_c^{(2)} (x) X_{T c}^{(2)} \right] 
\,, 
\end{eqnarray} 
where 
\begin{eqnarray} 
 X_{\rm eaten}^i 
 &=& 
\frac{1}{2} \left( 
\begin{array}{c|c} 
  \tau^i \otimes 1_{3\times 3}  &  \\ 
  \hline 
   & \tau^i 
\end{array}
\right)  \,, 
\qquad 
X^i_P 
= 
\frac{1}{2 \sqrt{3}}
\left( 
\begin{array}{c|c} 
  \tau^i \otimes 1_{3\times 3}  &  \\ 
  \hline 
   & -3 \cdot \tau^i 
\end{array}
\right)  \,, \qquad 
X_P
= 
\frac{1}{4 \sqrt{3}}
\left( 
\begin{array}{c|c} 
  1_{6 \times 6}  &  \\ 
  \hline 
   & -3\cdot 1_{2 \times 2} 
\end{array}
\right) 
\,, \nonumber \\ 
 X_{\theta a}^i 
 &=& 
\frac{1}{\sqrt{2}}  \left( 
\begin{array}{c|c} 
  \tau^i \otimes \lambda_a  &  \\ 
  \hline 
   & 0 
\end{array}
\right)  \,, 
\qquad 
X_{\theta_a} 
= 
\frac{1}{2 \sqrt{2}}
\left( 
\begin{array}{c|c} 
  1_{2\times 2} \otimes \lambda_a  &  \\ 
  \hline 
   & 0  
\end{array}
\right)  \,, \nonumber \\ 
X^{(1)i}_{T c} 
&=&  
\frac{1}{2}
\left( 
\begin{array}{c|c} 
 & \tau^i \otimes \xi_c    \\ 
  \hline 
\tau^i \otimes \xi_c^\dag  &  
\end{array}
\right) 
\,, \qquad 
X^{(2)i}_{T c}  
= 
\frac{1}{2}
\left( 
\begin{array}{c|c} 
 & - i \tau^i \otimes \xi_c    \\ 
  \hline 
i \tau^i \otimes \xi_c^\dag  &  
\end{array}
\right) 
\,, \nonumber \\ 
X^{(1)}_{T c}  
&=&  
\frac{1}{4}
\left( 
\begin{array}{c|c} 
 & 1_{2 \times 2} \otimes \xi_c    \\ 
  \hline 
1_{2 \times 2} \otimes \xi_c^\dag  &  
\end{array}
\right) 
\,, \qquad 
X^{(2)}_{T c} 
= 
\frac{1}{4}
\left( 
\begin{array}{c|c} 
 & - i \cdot 1_{2 \times 2} \otimes \xi_c    \\ 
  \hline 
i \cdot 1_{2 \times 2} \otimes \xi_c^\dag  &  
\end{array}
\right) 
\,, 
\end{eqnarray}
with $\xi_c$ being a three-dimensional unit vector in color space  
and the generators normalized as ${\rm Tr}[X^A X^B]=\delta^{AB}/2$. 
The color-triplet techni-pions 
$\{T_c^i, \bar{T}_c^i \}$ and $\{T_c, \bar{T}_c \}$ are respectively constructed from 
$\{ (T_c^{(1)})^i, (T_c^{(2)})^i\}$ and $\{ T_c^{(1)}, T_c^{(2)}  \}$ 
as 
\begin{eqnarray}  
T_c^i &=& \frac{ (T_c^{(1)})^i - i (T_c^{(2)})^i}{\sqrt{2}}  \,, \qquad 
\bar{T}_c^i = (T_c^i)^\dag 
\,, \nonumber \\  
T_c &=& \frac{ T_c^{(1)} - i T_c^{(2)} }{\sqrt{2}} 
\, , \qquad 
\bar{T}_c = (T_c)^\dag
\,. 
\end{eqnarray}

The covariant derivative $({\cal D}_\mu U)$ in Eq.(\ref{Lag:int}) now reads  
\begin{eqnarray} 
  {\cal D}_\mu U 
  &=& \partial_\mu U - i {\cal L}_\mu U + i U {\cal R}_\mu 
  \,, \nonumber \\ 
{\cal L}_\mu 
&=& 
2 g_W W_\mu^i X^i_{\rm eaten} + \frac{2}{\sqrt{3}} g_Y B_\mu X_P  + \sqrt{2} g_s  G_\mu^a X_{\theta_a}  
\,, \nonumber \\ 
{\cal R}_\mu 
&=&
2 g_Y B_\mu \left( X_{\rm eaten}^3 + \frac{1}{\sqrt{3}} X_P \right) + \sqrt{2} g_s G_\mu^a X_{\theta_a} 
\,. 
\end{eqnarray}
With this covariant derivative, one can easily see that the $|{\cal D}_\mu U|^2$ term in Eq.(\ref{Lag:int}) gives the $W$ boson mass,  
\begin{equation} 
  m_W^2 = \frac{1}{4 }g_W^2 (4 F_\pi^2) = \frac{1}{4} g_W^2 v_{\rm EW}^2  
\,, 
\end{equation}
as well as the $Z$ boson mass, where $v_{\rm EW} \simeq 246$ GeV.

In order to study the techni-pion LHC phenomenologies later, we shall next derive 
the techni-pion couplings to the SM particles.

\subsubsection{Couplings to the SM gauge bosons}

 The techni-pion couplings to two SM gauge bosons arise from the 
(covariantized) Wess-Zumino-Witten (WZW) term~\cite{Wess:1971yu} 
related to the non-Abelian $SU(8)_L \times SU(8)_R$ anomaly,  
\begin{equation} 
  S_{\rm WZW} [U, {\cal L}, {\cal R}] 
\,,  
\end{equation}
which includes the couplings as  
\begin{eqnarray} 
  S_{\rm WZW}[U, {\cal L}, {\cal R}] 
&\ni& 
  - \frac{N_{\rm TC}}{48\pi^2} \int_{M^4} \Bigg\{ 
  {\rm tr}[ (d {\cal L} {\cal L} + {\cal L} d {\cal L}) \alpha + (d {\cal R} {\cal R} + {\cal R} d {\cal R}) \beta ]
\nonumber\\ 
 && 
 \hspace{80pt}
+ i \, {\rm tr}[ d {\cal L} d U {\cal R} U^\dag - d {\cal R} d U^\dag {\cal L} U] 
 \Bigg\} 
 \nonumber \\ 
&=& 
- \frac{N_{\rm TC}}{12 \pi^2 F_\pi} \int_{M^4} 
{\rm tr} \left[ 
\left(3 d {\cal V} d {\cal V} + d {\cal A} d {\cal A}  \right) \pi 
+ {\cal O}(\pi^2) 
\right] 
\,, \label{pi-g-g}
\end{eqnarray}
where $M^4$ denote a four-dimensional Minkowski manifold 
and the things have been written in differential form, and 
\begin{equation} 
  \alpha = - i d U U^\dag 
  \,, \qquad 
  \beta = - i  U^\dag d U 
  \,. 
\end{equation}
The vector and axialvector fields, ${\cal V}$ and ${\cal A}$, are expressed in terms of $W^\pm, Z$, 
photon ($A$) and gluon ($G$) fields 
as follows: 
\begin{eqnarray} 
 {\cal V} 
 &\equiv & 
\frac{{\cal R} + {\cal L}}{2} 
=  
g_s G^a \Lambda^a + e Q_{\rm em} A + \frac{e}{2sc} \left( I_3 - 2 s^2 Q_{\rm em}  \right) Z 
+ \frac{e}{2 \sqrt{2}s} \left( W^+ I_+ + W^- I_- \right) 
\,, \nonumber \\ 
{\cal A} 
&\equiv & 
\frac{{\cal R}-{\cal L}}{2} =  
- \frac{e}{2sc} I_3 \, Z - \frac{e}{2 \sqrt{2} s} \left( W^+ I_+  +  W^- I_- \right) 
\,, 
\end{eqnarray}
where $s$ ($c^2=1-s^2$) denotes the standard weak mixing angle defined by $g_W=e/s$ and $g_Y=e/c$, and  
\begin{eqnarray} 
\Lambda_a &=& \sqrt{2} \, X_{\theta_a} \,, \qquad  
I_3 = 2 \, X_{\rm eaten}^3 \,  , \qquad 
Q_{\rm em} = I_3 + Y \,, \nonumber \\ 
Y&=& \frac{2}{\sqrt{3}} X_P  \,, \qquad 
I_+ = 2(X_{\rm eaten}^1 + i X_{\rm eaten}^2)   \,, \qquad 
I_- = (I_+)^\dag 
\,. 
\end{eqnarray}
 Substituting these expressions into the last line of Eq.(\ref{pi-g-g}) we  find the techni-pion couplings to two 
 gauge bosons.  To the neutral and colorless pion $P^0$ and color-octet pion $\theta_a$, for instance, we have 
\begin{eqnarray} 
S_{P^0 g g} 
&=&  - \frac{N_{\rm TC}}{16 \sqrt{3} \pi^2} \frac{g_s^2}{F_\pi} \int_{M^4}   P^0 d G^a d G^a 
\,, \nonumber \\ 
S_{P^0 \gamma \gamma} 
&=&  \frac{N_{\rm TC}}{12 \sqrt{3} \pi^2} \frac{e^2}{F_\pi} \int_{M^4}   P^0 d A d A 
\,, \nonumber \\ 
S_{P^0 Z \gamma} 
&=&  \frac{N_{\rm TC}}{6 \sqrt{3} \pi^2} \frac{e^2 s}{c F_\pi} \int_{M^4}   P^0 d Z d A 
\,, \nonumber \\ 
S_{P^0 Z Z} 
&=&  \frac{N_{\rm TC}}{12 \sqrt{3} \pi^2} \frac{e^2 s^2}{c^2 F_\pi} \int_{M^4}   P^0 d Z d Z 
\,, \nonumber \\ 
S_{P^0 W W} 
&=& 
0 
\,, \label{P0-gauge-int} 
\end{eqnarray} 
and 
\begin{eqnarray}
S_{\theta g g} 
&=&  -  \frac{N_{\rm TC}}{8 \sqrt{2} \pi^2} \frac{g_s^2}{F_\pi} \int_{M^4}  d_{abc} \theta^a d G^b d G^c 
\,, \qquad 
d_{abc} \equiv \frac{1}{4} {\rm Tr}[\lambda_a \{\lambda_b, \lambda_c \}]
\,, \nonumber \\ 
S_{\theta Z g} 
&=&  - \frac{N_{\rm TC}}{12 \sqrt{2} \pi^2} \frac{g_s es}{F_\pi c} \int_{M^4}  \theta_a d Z d G^a 
\,, \nonumber \\ 
S_{\theta Z \gamma} 
&=&  - \frac{N_{\rm TC}}{12 \sqrt{2} \pi^2} \frac{g_s e}{F_\pi } \int_{M^4}  \theta_a d A d G^a 
\,. \label{P8-gauge-int} 
\end{eqnarray} 
 Note that the $P^0$-$W$-$W$ coupling vanishes because the vertex $\propto {\rm tr}[X_P]=0$ which means the cancellation 
 between techni-quark and -lepton contributions.

\subsubsection{Couplings to the SM fermions}

As was noted at the beginning of this section, the Yukawa couplings between 
the techni-pions and SM fermions depend on models of ETC. 
 We shall here consider a typical ETC embedding the one-family techni-fermions and SM fermions  
 in a single multiplet. 
We assume that the ETC carries no SM charges and chiral 
techni-quarks $Q_{L,R}=(U, D)_{L,R}$ and -leptons $L_{L,R}=(N,E)_{L,R}$ are separately included in 
the ETC multiplets, ${\cal Q}_{L,R}=\{Q, q \}_{L,R}$ and 
${\cal L}_{L,R}=\{ L, l \}_{L,R}$, 
along with the SM quarks $q_{L,R}=(q_u, q_d)_{L,R}$ and leptons $l_{L,R}=(\nu, \ell)_{L,R}$.  
We focus only on flavor-diagonal couplings 
to avoid the FCNC problem. 
 Then the ETC gauge boson exchanges generically generate the 
induced four-fermion interactions at the scale $\Lambda_{\rm ETC}$ as 
\begin{eqnarray} 
  {\cal L}_{\rm ETC}^{\rm eff} 
  &=&-  \frac{1}{\Lambda_{\rm ETC}^2} \Bigg[  
\bar{\cal Q}_L^i \gamma_\mu (T^{\cal Q})_{ij} {\cal Q}_{L}^j 
\cdot \bar{\cal Q}_R \gamma^\mu (T^{\cal Q})_{kl} {\cal Q}_R^k 
\,, \nonumber \\ 
  && 
\hspace{30pt}    + 
    \bar{\cal L}_L^i \gamma_\mu (T^{\cal L})_{ij} {\cal L}_L^j 
\cdot \bar{\cal L}_R \gamma^\mu (T^{\cal L})_{kl} {\cal L}_R^k 
\Bigg] 
\,, 
\end{eqnarray}
 where $T^{\cal Q}$ and $T^{\cal L}$ denote the ETC generators corresponding to the ${\cal Q}$ and ${\cal L}$ multiplets, respectively. 
 Performing Fiertz rearrangement and picking up only the scalar $(S)$ and pseudo-scalar $(pS)$ channels, 
 we are thus left with 
\begin{eqnarray} 
  {\cal L}_{\rm ETC}^{S,pS} 
  &=& G_{\cal Q} \left( \bar{Q}U \bar{q} q_u - \bar{\cal Q} \gamma_5 U \bar{q} \gamma_5 q_u +  \cdots 
\right)   
  \nonumber \\ 
   &&
 + G_{\cal L} \left( 
\bar{L}E \bar{l} \ell - \bar{E} \gamma_5 E \bar{l} \gamma_5 \ell +  \cdots 
\right)   
\,, \label{4-fermi-ints}
\end{eqnarray}   
where $G_{{\cal Q},{\cal L}} \sim 1/\Lambda_{\rm ETC}^2$ which involves all the numerical factors arising from  
the Fiertz transformation on the Dirac spinors, ETC generators $T^{\cal Q}$ and $T^{\cal L}$.  
 The first terms in the first and second lines lead to the SM quark and lepton masses through the techni-quark 
and -lepton condensates, 
 \begin{equation} 
   m_{q} = - G_{\cal Q} \langle \bar{Q} U \rangle   
   \,, \qquad 
   m_l = - G_{\cal L} \langle \bar{L} E \rangle 
   \,. 
   \label{q-l-masses}
 \end{equation}

 We next consider the techni-pion couplings to techni-quarks and -leptons. 
They are completely determined by the low-energy theorem based on the Ward-Takahashi identities 
for the axialvector current $J_\mu^5$: 
\begin{equation} 
  \lim_{q_\mu \to 0}  q^\mu \int d^4 z e^{iqz} \langle 0| T J^5_\mu(z)  \bar{F}(x) F(0) |0 \rangle 
    = \delta_5 \langle 0 | T \bar{F}(x) F(0) |0 \rangle 
    \,, \label{WT:pi}
\end{equation} 
where $\delta_5 {\cal O}= [i Q_5, {\cal O}]$ denotes the infinitesimal chiral transformation under the chiral charge 
$Q_5= \int d^3 x J_0^5(x)$ associated with 
the axialvector current $J_\mu^5$. 
 The left hand side is saturated by the techni-pion pole: 
\begin{equation} 
  ({\rm L.H.S}) =  F_\pi \langle \pi(q=0) | T F(x)\bar{F}(0) | 0 \rangle 
  \,, 
\end{equation}
 where the techni-pion decay constant has been defined as 
 \begin{equation} 
   \langle 0| J_\mu^5(x) | \pi(q)\rangle = - i F_\pi q_\mu e^{-iqx} 
   \,. 
 \end{equation}
 Taking the Fourier transform of Eq.(\ref{WT:pi}) with respect to $p$, 
we find the amputated Yukawa vertex function $\chi_{\pi FF}(0,p)$:  
\begin{eqnarray} 
 \chi_{\pi FF}(0,p) &\equiv& S_F^{-1}(p) \cdot \langle \pi(q=0) | T F(x)\bar{F}(0) | 0 \rangle \cdot S^{-1}_F(p) 
 \nonumber \\
 &=& S_F^{-1}(p) \cdot  \left(\frac{1}{F_\pi} \delta_5 S_F( p) \right) \cdot S^{-1}_F(p) 
 = - \frac{1}{F_\pi} \delta_5 S_F^{-1}(p) 
  \,. \label{chi:P0FF}
\end{eqnarray}  
with $S_F(p)$ being the (full) $F$-fermion propagator.

   To be concrete,  consider the $P^0$ techni-pion. 
Then the chiral transformations for the techni-quark and -leptons 
are read off from Table~\ref{tab:TP} as  
$\delta_{P^0} Q(x) = -  \frac{i}{4 \sqrt{3}} \gamma_5 Q(x)$ and $\delta_{P^0} L(x) =  \frac{3i}{4 \sqrt{3}} \gamma_5 L(x)$, 
so that 
\begin{eqnarray}
\delta_{P^0} S_Q^{-1}(p) &=& - \frac{1}{4\sqrt{3}}\,  i \{ \gamma_5,  S_Q^{-1}(p) \} 
= - \frac{1}{2\sqrt{3}} \, i \gamma_5 \Sigma_Q(p^2) \cdot Z_Q(p^2) 
\,, \nonumber \\   
\delta_{P^0} S_L^{-1}(p) &=&  \quad \, \frac{3}{4\sqrt{3}}\, i \{ \gamma_5,  S_L^{-1}(p) \} 
= \quad \, \frac{3}{2\sqrt{3}} \, i \gamma_5 \Sigma_L(p^2) \cdot Z_L(p^2) 
\,, \label{P0:trans:SF}
\end{eqnarray}
 where we have parametrized the $F$-fermion propagator as $S_F(p) = [i Z_F(p^2) (\Sigma_F(p^2) - \slash \hspace{-5pt}p)]^{-1}$ 
with the mass function $\Sigma_F(p^2)$ and wavefunction renormalization $Z_F(p^2)$.  
Putting Eq.(\ref{P0:trans:SF}) into Eq.(\ref{chi:P0FF}) and  
 defining the renormalized Yukawa vertex function as 
\begin{equation} 
\chi^R_{P^0 FF}(0,p) \equiv Z_F^{-1}(p^2) \chi_{P^0 FF}(0,p)
\,,
\end{equation}  
we thus find 
\begin{eqnarray} 
  \chi^R_{P^0 QQ}(0,p) &=& 
\frac{i}{2\sqrt{3}} \frac{\gamma_5 \Sigma_Q(p^2)}{F_\pi} 
\,, \nonumber \\ 
  \chi^R_{P^0 LL}(0,p) &=& 
- \frac{3i}{2\sqrt{3}} \frac{\gamma_5 \Sigma_L(p^2)}{F_\pi} 
\,. \label{BSamps}
\end{eqnarray}

 \begin{figure}

\begin{center} 
\includegraphics[scale=0.7]{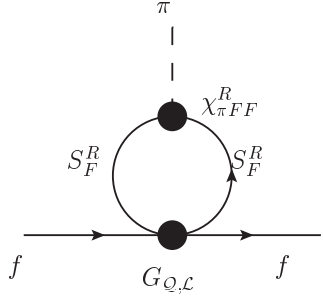}
\end{center}
\caption{The diagram yielding the techni-pion Yukawa vertex amplitude 
 to the SM quarks and leptons such as Eq.(\ref{iM}),  
involving the ETC-induced four-fermion vertices in Eq.(\ref{4-fermi-ints}) and the techni-pion 
Yukawa vertices to the SM fermions in Eq.(\ref{BSamps}). }   
\label{TP-yukawa}
\end{figure}

 Now that we have obtained the Yukawa vertex functions in Eq.(\ref{BSamps}) and specified 
the ETC-induced four-fermion terms in Eq.(\ref{4-fermi-ints}),  
it is straightforward to calculate the techni-pion $(P^0)$ Yukawa coupling to the SM quarks and leptons 
just by evaluating the following amplitude (See Fig.~\ref{TP-yukawa}): 
\begin{eqnarray} 
  i {\cal M}(P^0(0), f(p),f(p)) 
  &=& - C_f \cdot \frac{i G_{{\cal Q},{\cal L}}}{2 \sqrt{3}F_\pi} {\rm Tr} \int \frac{d^4 k}{(2 \pi)^4} \left[ 
S_F^R(k) \gamma_5 S_F^R(k) \gamma_5 \Sigma(k^2) \right] \cdot 
 \bar{u}_f(p) \gamma_5 u_f(p)
\,, \label{iM}
\end{eqnarray}
  where $C_f=(1,-3)$ for $f=(q,l)$, $u_f(p)$ is the wavefunction of $f$-fermion and   
$S_F^R(p)$ is the renormalized propagator for the $F$-fermion defined as $S_F^R(p)=[i(\Sigma_F(p^2) - \slash \hspace{-5pt}p)]^{-1}$. 
 This amplitude is rewritten as 
\begin{eqnarray} 
   i {\cal M}(P^0(0), f(p),f(p)) 
  &=& - C_f \cdot \frac{G_{{\cal Q},{\cal L}}}{2 \sqrt{3}F_\pi} {\rm Tr} \int \frac{d^4 k}{(2 \pi)^4} \left[ 
  S_F^R(k) \right]
  \cdot 
 \bar{u}_f(p) \gamma_5 u_f(p)
\nonumber \\ 
 &=&
 C_f \cdot \frac{G_{{\cal Q},{\cal L}}}{2 \sqrt{3}F_\pi} 
 \cdot 
 \langle \bar{F}F \rangle \cdot 
 \bar{u}_f(p) \gamma_5 u_f(p) 
 \,. 
 \end{eqnarray} 
 Noting Eq.(\ref{q-l-masses}), we find 
 \begin{eqnarray} 
  i {\cal M}(P^0,q(p),q(p)) 
  &=& 
  \frac{m_q}{2\sqrt{3} F_\pi} \cdot 
 \bar{u}_q(p) \gamma_5 u_q(p) 
 \,, \nonumber \\ 
   i {\cal M}(P^0,l(p),l(p)) 
  &=& 
  - \frac{3m_l}{2\sqrt{3} F_\pi} \cdot 
 \bar{u}_l(p) \gamma_5 u_l(p) 
\,. 
\end{eqnarray}   
 These matrix elements imply the Yukawa coupling terms: 
\begin{equation} 
 {\cal L}_{P^0 ff} 
   =
- \frac{i}{ 2 \sqrt{3} F_\pi} P^0 \left[ \sum_q m_q \bar{q} \gamma_5 q - 3 \sum_l m_l \bar{l} \gamma_5 l \right]
\,. \\  \label{P0-yukawa-int} 
\end{equation}

One can easily derive  similar formulas for other techni-pions.  
For instance, it turns out that the $\theta_a$-$f$-$f$ coupling takes the form 
~\footnote{
A set of more general Yukawa coupling terms was discussed in Ref.~\cite{Casalbuoni:1992nw}. }    
\begin{eqnarray} 
{\cal L}_{\theta ff} 
= 
- \frac{\sqrt{2} i}{F_\pi} \theta_a \left[ \sum_q m_q \bar{q} \gamma_5 \left( \frac{\lambda_a}{2} \right) q \right]
\,.  
\label{theta-yukawa-int} 
\end{eqnarray}

\subsection{Techni-pion masses}

In this subsection we shall calculate the techni-pion masses in the one-family model 
embedded in the walking TC. 
The masses of techni-pions arise as explicit breaking effects of the full chiral $SU(8)_L \times SU(8)_R$ 
symmetry associated with the chiral 
transformation yielding the chiral currents as listed in Table~\ref{tab:TP}.  
There are two sources giving such explicit breaking effects: one is from the SM gauge interactions, while 
the other from ETC-induced four-fermion interactions.

\subsubsection{Electroweak-origin mass}

\begin{figure}
\begin{center}
   \includegraphics[scale=0.6]{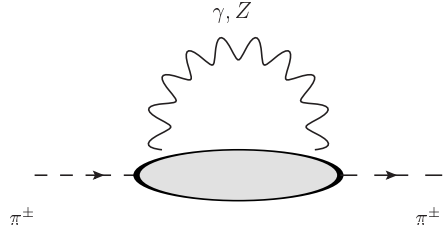}
\caption{ 
The one-photon and $Z$ boson exchange graphs contributing to the 
masses of the electrically charged pions $\pi^\pm= 
\{ \theta_a^\pm, T_c^\pm (\bar{T}_c^\pm), P^\pm\}$. 
\label{pimass-diff}
}
\end{center} 
 \end{figure}

The EW radiative corrections give rise to masses for the charged techni-pions, $\pi^\pm= 
\{ \theta_a^\pm, T_c^\pm (\bar{T}_c^\pm), P^\pm\}$, 
analogously to electromagnetic corrections to the charged pion mass in QCD. 
The charged pion mass-squared $\Delta m_{\pi^{\pm}}^2$ can be estimated by 
taking into account one-photon and -$Z$ boson exchanges as illustrated in Fig.~\ref{pimass-diff}:  
\begin{eqnarray} 
  \Delta m_{\pi^{\pm}}^2 
&= & 
(\Delta m_{\pi^\pm}^2)_\gamma + (\Delta m_{\pi^\pm}^2)_Z 
\,, \nonumber \\ 
(\Delta m_{\pi^\pm}^2)_\gamma
&=&
- \frac{i}{2} e^2 \int d^4 x  D_{\mu\nu}^{(\gamma)}(x)  
\langle \pi^+ | T J_{\rm em}^\mu(x) J_{\rm em}^\nu(0)  | \pi^+ \rangle
\,, \nonumber \\ 
(\Delta m_{\pi^\pm}^2)_Z 
&=&
- \frac{i}{2} \frac{e^2}{4 s^2c^2} \int d^4 x D_{\mu\nu}^{(Z)}(x) 
\langle \pi^+ | T J_Z^\mu(x) J_Z^\nu(0)  | \pi^+ \rangle
\,, 
  \,, \label{mass:diff}
\end{eqnarray}
where $D_{\mu\nu}^{(\gamma, Z)} $ denote the photon and $Z$ boson propagators, $s^2(=1-c^2)$ stands for the usual 
weak mixing angle and $J^\mu_{{\rm em}, Z}$ are the electromagnetic and $Z$ boson currents composed of techni-fermions $(Q_c, L)$
defined by  
\begin{eqnarray} 
{\cal L}_{\gamma FF, ZFF} 
&=& 
e J_{\rm em}^\mu A_\mu + \frac{e}{2sc} J_Z^\mu Z_\mu 
\,, \nonumber\\  
J_{\rm em}^\mu &=& \bar{Q}_c \gamma^\mu 
\left( 
\begin{array}{cc} 
 2/3 & 0 \\ 
 0 & -1/3   
\end{array}
\right)
 Q_c 
+ \bar{L} \gamma^\mu 
\left( 
\begin{array}{cc} 
 0 & 0 \\ 
 0 & -1   
\end{array}
\right)
L 
\,, \nonumber \\ 
J_Z^{\mu}  
&=& 
(c^2-s^2) \left\{ \bar{Q}_c \gamma^\mu \tau^3 Q_c  + \bar{L} \gamma^\mu \tau^3 L \right\} 
- 2 s^2 \left\{  \frac{1}{6} \bar{Q}_c \gamma^\mu Q_c - \frac{1}{2} \bar{L} \gamma^\mu L  \right\} 
- \left\{ \bar{Q}_c \gamma^\mu \gamma_5 \tau^3 Q_c  + \bar{L} \gamma^\mu \gamma_5 \tau^3 L \right\} 
. 
\end{eqnarray}
By using the reduction formula together with the partially conserved axialvector current (PCAC): $\partial_\mu J^\mu_{\pi^\pm}(x) 
= \sqrt{2} F_\pi m_\pi^2 \pi^\pm(x)$, the current algebra technique allows us to 
rewrite Eq.(\ref{mass:diff}) in terms of the vector and axialvector current correlators.

To the colorless-electrically charged techni-pions $P^\pm$ coupled to the associated currents $J_{P^\pm}^\mu = \frac{1}{2 \sqrt{3}} 
  [ \bar{Q}_c \gamma^\mu \gamma_5 \tau^\pm Q_c - 3 \bar{L} \gamma^\mu \gamma_5 \tau^\pm L]$, 
we find 
\begin{eqnarray} 
  (\Delta m_{P^\pm}^2)_\gamma
  &=& 
  \frac{\alpha_{\rm EM}}{16 \pi F_\pi^2} \int_0^{\infty} 
  d Q^2  \left[ 
  \Pi_{V-A}^{Q}(Q^2) + 9 \, \Pi_{V-A}^L(Q^2)
\right]
\,, \nonumber \\ 
  (\Delta m_{P^\pm}^2)_Z
  &=& 
  - \frac{\alpha_{\rm EM}}{16 \pi F_\pi^2} \int_0^{\infty} 
  \frac{d Q^2\,  Q^2}{m_Z^2+Q^2} \left[ 
  \Pi_{V-A}^{Q}(Q^2) + 9 \, \Pi_{V-A}^L(Q^2)
\right]
\,,\label{deltaMP:pm}
\end{eqnarray}  
where $\alpha_{\rm EM}= e^2/(4 \pi)$ and 
$Q^2=-p^2$ denotes Euclidean momentum-squared and $\Pi_{V-A}^F(Q^2) \equiv \Pi_V^F(Q^2) - \Pi_A^F(Q^2)$  
($F= Q_c, L$) with $\Pi^F_{V(A)}$ being vector (axialvector) current correlator defined as 
\begin{eqnarray} 
  i \int d^4 x e^{- ipx} \langle 0| T \left( \bar{F}(x) \gamma^\mu \tau^a F (x) \bar{F}(0) \gamma^\nu \tau^b F (0)  \right) |0 \rangle 
  &=& \left( \frac{p^\mu p^\nu}{p^2} - g^{\mu\nu}   \right) \delta^{ab} \Pi_V^F(Q^2) 
  \,, \\ 
   i \int d^4 x e^{- ipx} \langle 0 |T \left( \bar{F}(x) \gamma^\mu \gamma_5 \tau^a F (x) \bar{F}(0) \gamma^\nu \gamma_5 \tau^b F (0)  \right) |0 \rangle 
  &=& \left( \frac{p^\mu p^\nu}{p^2} - g^{\mu\nu}   \right) \delta^{ab} \Pi_A^F(Q^2) 
  \,. 
\end{eqnarray}
  Note the relative sign between the photon and $Z$ boson contributions in Eq.(\ref{deltaMP:pm}), 
which give the dramatic cancellation in a way similar to the collective symmetry breaking in the little Higgs, such that the total contribution becomes  
\begin{equation} 
   \Delta m_{P^\pm}^2 
  = 
  \frac{\alpha_{\rm EM}}{16 \pi F_\pi^2} \int_0^{\infty} \, d Q^2 \,  
  \frac{m_Z^2}{m_Z^2+Q^2} \left[ 
  \Pi_{V-A}^{Q}(Q^2) + 9 \, \Pi_{V-A}^L(Q^2)
\right]
\,. \label{mass:diff:tot}
\end{equation} 
It is easy to derive similar formulas for the other charged techni-pions as well.

The right hand side of Eq.(\ref{mass:diff:tot}) can be split into two terms: 
\begin{eqnarray} 
  \int_0^\infty 
  d Q^2  \frac{m_Z^2}{m_Z^2+Q^2}  \Pi_{V-A}^F(Q^2) =
 \int_0^{\Lambda_\chi^2}   
  d Q^2  \frac{m_Z^2}{m_Z^2+Q^2}  \Pi_{V-A}^F(Q^2) 
  + \int_{\Lambda_\chi^2}^{\Lambda_{\rm TC}^2 (\to \infty)}  
  d Q^2  \frac{m_Z^2}{m_Z^2+Q^2}  \Pi_{V-A}^F(Q^2) 
  \,, \label{split}
\end{eqnarray}
where $\Lambda_\chi \simeq \frac{4 \pi F_\pi}{\sqrt{N_{\rm TF}}}$~\footnote{
For the presence of the number of fermions in the chiral symmetry breaking scale $\Lambda_\chi$, 
see Refs.~\cite{Soldate:1989fh}. } above which scale ($Q^2 > \Lambda^2_\chi$) 
the operator product expansion  for $\Pi^F_{V,A}(Q^2)$ is assumed to be valid. 
The contributions in the infrared region $(Q^2 < \Lambda_{\chi}^2)$ 
can be computed using the current algebra~\cite{Eichten:1979ah} or chiral perturbation~\cite{Peskin:1980gc,Chadha:1981yt}, 
so that the first term in Eq.(\ref{split}) yields 
\begin{equation} 
(\Delta m_{P^\pm}^2)_{Q^2 < \Lambda_\chi^2} = \frac{3 \alpha_{\rm EM}}{4 \pi} m_Z^2 
\log \frac{\Lambda^2_{\chi}}{m_Z^2}  
\simeq (9\, {\rm GeV})^2 
\,, \label{mass:diff:IR}
\end{equation} 
for $F_\pi=123$ GeV. 
On the other hand, 
the ultraviolet region $(Q^2 > \Lambda_\chi^2)$ we may use the operator product expansion,  
\begin{equation} 
\Pi_{V-A}^F(Q^2) \stackrel{Q^2 > \Lambda_\chi^2}{\simeq} 
\frac{4(N_{\rm TC}^2 -1)}{N_{\rm TC}^2}  \left(\frac{Q^2}{\mu^2} \right)^{\gamma_m} 
\frac{\alpha(\mu) \langle \bar{F}F  \rangle^2_\mu}{Q^4} 
\,, \label{OPE}
\end{equation}  
with a  renormalization scale $\mu$. 
  Taking into account $\langle \bar{L}L \rangle = 1/3 \langle \bar{Q}Q \rangle$ and letting $\langle \bar{L}L \rangle$ be 
$\langle \bar{F}F \rangle$, 
we thus find 
\begin{eqnarray} 
\int_{\Lambda_\chi^2}^{\Lambda_{\rm TC}^2 (\to \infty)}  
  d Q^2  \frac{m_Z^2}{m_Z^2+Q^2}  \left[ 
  \Pi_{V-A}^{Q}(Q^2) + 9 \, \Pi_{V-A}^L(Q^2)
\right]
  \simeq 
  \frac{48 \pi \langle \bar{F}F \rangle^2_{\Lambda_\chi}}{N_{\rm TC} \Lambda_\chi^2} \log \left( 1 + \frac{m_Z^2}{\Lambda_\chi^2} \right)
  \,, \label{mass:diff:2}
\end{eqnarray} 
where we used $\alpha \simeq \alpha_c = \pi/3(C_2(F))=\frac{2\pi N_{\rm TC}}{3(N_{\rm TC}^2-1)}$~\cite{Hashimoto:2010nw}.

  In order to make the right hand side of Eq.(\ref{mass:diff:2}) more explicit, 
we shall evaluate the chiral condensate $\langle \bar{F}F  \rangle$,   
\begin{equation} 
    \langle \bar{F}F \rangle_{\Lambda_{\rm TC}} 
    = - \frac{N_{\rm TC}}{4\pi^2} m_F^3 \int_0^{\Lambda_{\rm TC}^2/m_F^2 \to \infty} 
    d x \frac{x \Sigma(x)}{x + \Sigma^2(x)} 
    \,,  
\end{equation} 
where $\Sigma(x)=\Sigma(-p^2)/m_F$ denotes the mass function of techni-fermion normalized as $\Sigma(1)=1$. 
At the dynamical techni-fermion mass scale $m_F$, the chiral condensate may be defined as 
\begin{equation} 
  \langle \bar{F}F \rangle_{m_F} 
    \equiv
   - \kappa_c \frac{N_{\rm TC}}{4\pi^2} m_F^3 
   \,, \label{condensate}
\end{equation} 
where $\kappa_c$ is an overall coefficient to be determined once the nonperturbative calculation is done. 
 Using the scaling law 
 \begin{equation} 
  \langle \bar{F}F \rangle_{\Lambda_{\rm TC}} \simeq \left( \frac{\Lambda_{\rm TC}}{\mu} \right)^{\gamma_m} 
  \langle \bar{F}F \rangle_{\mu}  
  \,, \qquad 
  \gamma_m \simeq 1 
  \,. \label{scaling} 
 \end{equation} 
Note that the dynamical mass $m_F$ can in general be related to the techni-pion decay constant $F_\pi$ as    
\begin{eqnarray} 
 F_\pi^2 & \equiv & 
 \kappa_F^2 \frac{N_{\rm TC}}{4 \pi^2} m_F^2 
\,, \label{Fpi:mF}
\end{eqnarray} 
with the overall coefficient $\kappa_F$ to be fixed by the straightforward calculation. 
 Using Eqs.(\ref{condensate}) and (\ref{Fpi:mF}) together with Eq.(\ref{scaling})
we thus express the chiral condensate renormalized at $\Lambda_\chi$ in terms of $F_\pi$: 
\begin{equation} 
   \langle \bar{F}F \rangle_{\Lambda_\chi} 
= - \left( \frac{\kappa_c}{\kappa_F^2} \right) \Lambda_\chi F_\pi^2 
   \,. \label{condensate:chi}
\end{equation}

Putting Eq.(\ref{condensate:chi}) into Eq.(\ref{mass:diff:2}) and taking into account Eq.(\ref{mass:diff:IR}),  
we arrive at a concise formula, 
\begin{eqnarray} 
    \Delta m_{P^{\pm}}^2 \simeq 
\frac{3 \alpha_{\rm EM}(\Lambda_\chi)}{4 \pi} m_Z^2 
\log \frac{\Lambda^2_{\chi}}{m_Z^2}  
+ 
\frac{3 \alpha_{\rm EM}(\Lambda_\chi)}{N_{\rm TC}} \left( \frac{\kappa_c}{\kappa_F^2} \right)^2 
F_\pi^2 \log \left( 1 + \frac{m_Z^2}{\Lambda_\chi^2} \right) 
\,. \label{mass:diff:concise}
\end{eqnarray}
  Note that the second term from the ultraviolet region $(Q^2> \Lambda_\chi^2)$ is almost negligible since 
$(m_Z/\Lambda_\chi)^2 \sim 10^{-3} \times N_{\rm TF}$.  
Thus the EW corrections to the charged techni-pion mass in the walking TC are dominated by 
the infrared contributions of order of a few GeV (Eq.(\ref{mass:diff:IR})), 
to be negligible compared to another source from ETC as will be discussed later.

\subsubsection{QCD-origin mass}

The QCD-gluon exchanges give masses to the colored techni-pions, 
$\theta_a^i$, $\theta_a$ and $T_c^i$ ($\bar{T}_c^i$). 
Those corrections can be estimated in a way similar to the photon contribution to the 
charged pion mass discussed above, simply by scaling $(\Delta m_{\pi^\pm}^2)_\gamma$ in Eq.(\ref{mass:diff}):  
\begin{equation} 
\frac{  \Delta m_{3,8}^2 }{ (\Delta m_{\pi^{\pm}}^2)_\gamma} 
= 
C_2(R) \frac{\alpha_s(\Lambda_\chi)}{\alpha_{\rm EM}(\Lambda_\chi)}
\,, \label{scaling:mpi}
  \end{equation} 
  where $C_2(R)=\frac{4}{3}(3)$ for color-triplets (-octets). 
The photon exchange contribution $(\Delta m_{\pi^{\pm}}^2)_\gamma$ 
is decomposed into two parts, infrared and ultraviolet terms, 
in the same way as done in Eq.(\ref{split}). 
The ultraviolet term then turns out to be highly dominant due to the large logarithmic enhancement coming from 
the slow damping behavior of $\Pi_{V-A}(Q^2)$ in the walking TC, $\Pi_{V-A}^F(Q^2) \sim \langle \bar{F}F \rangle^2/Q^2$ 
(See Eq.(\ref{OPE})): 
\begin{eqnarray} 
( \Delta m_{\pi^\pm}^2 )_\gamma 
 &\simeq & \frac{ 9 \alpha_{\rm EM}}{8 \pi F_\pi^2}  \int_{\Lambda_\chi^2}^{\Lambda_{\rm TC}^2 (\to \infty)} 
 \Pi_{V-A}^F(Q^2) 
 \nonumber \\ 
 &\simeq& 
 \frac{3 \alpha_{\rm EM}}{N_{\rm TC}}  \left( \frac{\kappa_c}{\kappa_F^2} \right)^2 
F_\pi^2 \log \frac{\Lambda^2_{\rm TC}}{\Lambda_\chi^2}
\,, 
\end{eqnarray}
where we used Eq.(\ref{condensate:chi}) and put $\alpha(\Lambda_\chi) = \alpha_c=\pi/(3 C_2(F))$.  
The colored techni-pion masses are thus estimated for triplets and octets as follows: 
\begin{eqnarray}     
\Delta m_3 &\simeq& 299 \, (358) \, {\rm GeV} \sqrt{\frac{3}{N_{\rm TC}}} \left(  \frac{\kappa_c}{4.0} \right)  \left( \frac{1.4}{\kappa_F} \right)^2
\nonumber \\ 
\Delta m_8 &\simeq& 449 \, (537)  \, {\rm GeV}  \sqrt{\frac{3}{N_{\rm TC}}} \left(  \frac{\kappa_c}{4.0} \right)  \left( \frac{1.4}{\kappa_F} \right)^2
\,, \label{m3m8}
\end{eqnarray} 
 for $\Lambda_{\rm TC} \simeq 10^3 (10^4)$ TeV. 
Here we have used $\alpha_s(\Lambda_\chi) \simeq 0.1$, $F_\pi = 123$ GeV and taken the values of $\kappa_c$ and $\kappa_F$ 
 from the recent result based on the ladder Schwinger-Dyson analysis~\cite{Hashimoto:2010nw}.

The estimated numbers in Eq.(\ref{m3m8}) are compared with those 
based on a naive scale-up version of QCD~\cite{Farhi:1980xs}, which are obtained by replacing $(\Delta m_{\pi^{\pm}}^2)_\gamma$ 
with $(\Delta m_{\pi^{\pm}}^2)_{\rm QCD} \simeq (35 \, {\rm MeV})^2$ and supplying the scaling factor $(F_\pi/f_\pi)^2 \simeq (1323)^2$ 
in the right hand side of Eq.(\ref{scaling:mpi}): 
\begin{equation}
\textrm{QCD-scale up} \, : 
\qquad \qquad 
\Delta m_3 \simeq 193 \, {\rm GeV} \sqrt{\frac{3}{N_{\rm TC}}}  , \qquad 
\Delta m_8 \simeq  290   \, {\rm GeV} \sqrt{\frac{3}{N_{\rm TC}}} 
\,. 
\end{equation} 
 This implies that the masses are enhanced by about 50\% (85\%) for $\Lambda_{\rm TC}=10^3(10^4)$ TeV due to the walking dynamics yielding 
the slow damping behavior of $\Pi_{V-A}(Q^2)$ in the high momentum region, in accord with a 
recent discussion in Ref.~\cite{Harada:2005ru}.

\subsubsection{ETC-origin mass}

The techni-pions $P^{i, 0}$ ($i=1,2,3$) associated with the currents generated by the separate chiral rotations between 
techni-quarks and -leptons may acquire the masses by ETC-induced four-fermion interactions as in Eq.(\ref{4-fermi-ints}), 
\begin{equation} 
{\cal L}_{\rm 4-fermi}^{\rm ETC} (\Lambda_{\rm ETC})
= \frac{1}{\Lambda_{\rm ETC}^2}
\left[ 
 \bar{Q}Q \bar{L}L -  \bar{Q} \gamma_5 \sigma^a Q \bar{L} \gamma_5 \sigma^a L 
\right]\,, 
\end{equation}
which is SM gauge-invariant but breaks the full chiral symmetry into the separate chiral symmetries associated with 
the techni-quarks and -leptons. 
The masses are then calculated in a way similar to the gauge boson exchange contributions to the charged pion masses in Eq.(\ref{mass:diff}) 
with use of the reduction formula, current algebra and associated PCAC relations 
$\partial_\mu J^{\mu}_{P^{i,0}}(x) = F_\pi m_P^2 P^{i,0}(x) $: 
\begin{eqnarray} 
( \Delta m_{P^{i, 0}}^2 )_{\rm ETC}
 &=& - \langle P^{i, 0} |  
 {\cal L}_{\rm 4-fermi}^{\rm ETC}(\Lambda_{\rm ETC})  
| P^{i, 0} \rangle 
\nonumber \\ 
&=& \frac{1}{F_\pi^2}
\langle 0| 
 [{\bf Q}_{P^{i,0}} ,[{\bf Q}_{P^{i,0}}, {\cal L}_{\rm 4-fermi}^{\rm ETC}(\Lambda_{\rm ETC}) ] ] 
|0 \rangle 
\,,
\end{eqnarray}
where ${\bf Q}_{P^{i,0}}$ denote the chiral charges defined as  
${\bf Q}_{P^{i,0}} = \int d^3 x J_{P^{i,0}}^0(x)$.  
 For each techni-pion, we thus find 
\begin{eqnarray} 
 ( \Delta m_{P^0}^2 )_{\rm ETC} 
&=& 
\frac{40}{48} \frac{ \langle 0| (\bar{Q} Q  \bar{L} L)_{\Lambda_{\rm ETC}} |0\rangle }{F_\pi^2 \Lambda_{\rm ETC}^2} 
= \frac{5}{2} \frac{ \langle 0| (\bar{F}F)_{\Lambda_{\rm ETC}} |0\rangle^2 }{F_\pi^2 \Lambda_{\rm ETC}^2} 
\,, \nonumber \\ 
 ( \Delta m_{P^i}^2 )_{\rm ETC} 
&=& 
\frac{16}{12} \frac{ \langle 0| (\bar{Q} Q  \bar{L} L)_{\Lambda_{\rm ETC}} |0\rangle }{F_\pi^2 \Lambda_{\rm ETC}^2} 
= 4 \frac{ \langle 0| (\bar{F}F)_{\Lambda_{\rm ETC}} |0\rangle^2 }{F_\pi^2 \Lambda_{\rm ETC}^2} 
\, . 
\end{eqnarray} 
Using Eqs.(\ref{scaling}) and (\ref{condensate:chi}),  
we arrive at 
\begin{eqnarray} 
\Delta m_{P^0}^{\rm ETC} 
&=& 
\sqrt{\frac{5}{2}} \left( \frac{\kappa_c}{\kappa_F^2} \right)  F_\pi \simeq 397 \, {\rm GeV} \left( \frac{\kappa_c}{4.0} \right) 
\left( \frac{1.4}{\kappa_F} \right)^2
\,, \nonumber \\ 
\Delta m_{P^i}^{\rm ETC} 
&=& 
 2 \left( \frac{\kappa_c}{\kappa_F^2} \right) F_\pi  \simeq 502 \, {\rm GeV} \left( \frac{\kappa_c}{4.0} \right) 
\left( \frac{1.4}{\kappa_F} \right)^2
\,, 
\end{eqnarray} 
where in the last expressions we have quoted the values of $\kappa_c$ and $\kappa_F$ from Ref.~\cite{Hashimoto:2010nw}. 
It is remarkable to note that since the $P^\pm$ mass becomes larger than the top quark mass, 
the current experimental limits on charged Higgs bosons~\cite{Searches:2001ac} 
are inapplicable to the walking $P^\pm$, where the limits are set based on the top quark decays to the charged Higgs. 
 A new proposal to constrain the walking $P^\pm$ is to be explored in the future.

To summarize, all the estimated masses that have been discussed so far are displayed in Table~\ref{tab:TP}.

\section{The LHC signatures of one-family walking techni-pions} 
 \label{LHC}

In this section we shall discuss the LHC signatures of the one-family walking techni-pions, 
especially focusing on neutral isosinglet scalars ($P^0$ and $\theta_a$), 
in comparison with the SM Higgs.

\subsection{Isosinglet-colorless techni-pion $P^0$}

 From Eq.(\ref{P0-gauge-int}) we compute the $P^0$ decay widths to the SM gauge boson pairs 
 to get 
\begin{eqnarray} 
\Gamma (P^0 \to gg) 
&=&
\frac{N_{\rm TC}^2 \alpha_s^2 G_F m^3_{P^0}}{12 \sqrt{2} \pi^3} 
\,, \nonumber \\ 
\Gamma (P^0 \to \gamma\gamma) 
&=&
\frac{N_{\rm TC}^2 \alpha_{\rm EM}^2 G_F m^3_{P^0}}{54 \sqrt{2} \pi^3} 
\,, \nonumber \\ 
\Gamma (P^0 \to Z \gamma) 
&=&
\frac{N_{\rm TC}^2 \alpha_{\rm EM}^2 G_F m^3_{P^0}s^2}{27 \sqrt{2} \pi^3 c^2} 
\left( 1- \frac{m_Z^2}{m_{P_0}^2} \right)^3 
\,, \nonumber \\ 
\Gamma (P^0 \to Z Z) 
&=&
\frac{N_{\rm TC}^2 \alpha_{\rm EM}^2 G_F m^3_{P^0}s^4}{54 \sqrt{2} \pi^3 c^4} 
\left( 1- \frac{4 m_Z^2}{m_{P_0}^2} \right)^{3/2} 
\,, \nonumber \\ 
\Gamma (P^0 \to WW) 
&=& 
0 \,, 
\end{eqnarray} 
where use has been made of $F_\pi=v_{\rm EW}/2$ and $1/v_{\rm EW}^2=\sqrt{2} G_F$ with $G_F$ being the Fermi constant. 
 Similarly from Eq.(\ref{P0-yukawa-int}), we also calculate the decay rates to the SM fermion pairs to find   
\begin{equation} 
 \Gamma(P^0 \to f \bar{f}) = 
A_f\cdot \frac{G_F m_{P^0} m_f^2}{4 \sqrt{2} \pi} \left( 1 - \frac{4 m_f^2}{m_{P^0}^2} \right)^{1/2} 
\,, 
\end{equation}
where $A_f=1(3)$ for quarks (leptons). 
The $P^0$ decay properties are summarized in Table~\ref{tab:P0-signals}.

\begin{table}
\begin{tabular}{|c||cccccc|} 
\hline 
\hspace{15pt} $N_{\rm TC}$ \hspace{15pt} 
& \hspace{15pt} $\Gamma_{\rm tot}^{P^0}$ [GeV]  \hspace{15pt} 
&\hspace{15pt} ${\rm BR}_{gg}$ \hspace{15pt} 
&\hspace{15pt} ${\rm BR}_{\tau^+ \tau^-}$ \hspace{15pt} 
&\hspace{15pt} ${\rm BR}_{t \bar{t}}$ \hspace{15pt} & &  \\
 \hline \hline 
3 & 4.0 & $4.3 \times 10^{-2}$ & $6.1 \times 10^{-4}$ & $9.5 \times 10^{-1}$  & &  \\ 
4 & 4.2 & $7.4 \times 10^{-2}$ & $5.9 \times 10^{-4}$ & $9.2 \times 10^{-1}$  & & \\
\hline 
 & \hspace{15pt} $r_{\rm GF}$  \hspace{15pt} 
&\hspace{15pt} $r_{\rm BR}^{gg}$ \hspace{15pt} 
&\hspace{15pt} $r_{\rm BR}^{\tau^+\tau^-}$ \hspace{15pt} 
&\hspace{15pt} $r_{\rm BR}^{t\bar{t}}$ \hspace{15pt}
&\hspace{15pt} $R_{\tau^+ \tau^-}$ \hspace{15pt}
&\hspace{15pt} $R_{t\bar{t}}$ \hspace{15pt} \\ 
 \hline \hline 
3 & 5.0 & 35 & 21 & 6.5 & 96 & 30 \\ 
4 & 9.0 & 60 & 20 & 6.3 & 165 & 52   \\
\hline 
\end{tabular}
\caption{
The $P^0$ total width, relevant branching fraction and numbers regarding the LHC signatures at 397 GeV. 
Here we have defined: 
$r_{\rm GF}\equiv \sigma_{\rm GF}^{P^0}/\sigma_{\rm GF}^{h_{\rm SM}}$ with the gluon fusion production cross section at $\sqrt{s}=$ 7 TeV 
$\sigma_{\rm GF}$; $r_{\rm BR}^X \equiv {\rm BR}(P^0 \to X)/{\rm BR}(h_{\rm SM} \to X)$; $R_X \equiv r_{\rm GF} \times r_{\rm BR}^X$. 
The branching ratios and 7 TeV LHC production cross section for the SM Higgs are taken from Ref.~\cite{Dittmaier:2011ti}.     
}
\label{tab:P0-signals} 
\end{table}

 Of interest is that the $P^0$ decays to $W$ and $Z$ boson pairs are highly suppressed,   
as was noted in Ref.~\cite{Chivukula:2011ue}, due to the techni-quark and lepton cancellation 
in loops. 
The $P^0$ is thus almost completely gaugephobic to be definitely distinguishable from 
the SM Higgs at the LHC. 
 Besides the obvious $WW$ mode, one can indeed check the gaugephobicity also to $ZZ$ mode by evaluating 
a ratio of the $P^0 \to ZZ$ decay width to the corresponding quantity of the SM Higgs 
which roughly scales like  
\begin{equation}
\frac{\Gamma(P^0 \to ZZ)}{\Gamma(h_{\rm SM} \to ZZ)}
  \approx 
 \left(  \frac{N_{\rm TC}}{3} \right)^2 
 \left( \frac{\sqrt{2}\alpha_{\rm EM} }{\sqrt{3} \pi} \frac{s^2}{c^2} \right)^2 
 \sim  10^{-7} \times \left(  \frac{N_{\rm TC}}{3} \right)^2 
 \,. \label{P0WW/ZZ}
\end{equation}  
Thus the $P^0$ signals through decays to the weak gauge bosons are to be almost invisible at the LHC.

\begin{figure}

\begin{center} 
\includegraphics[scale=0.6]{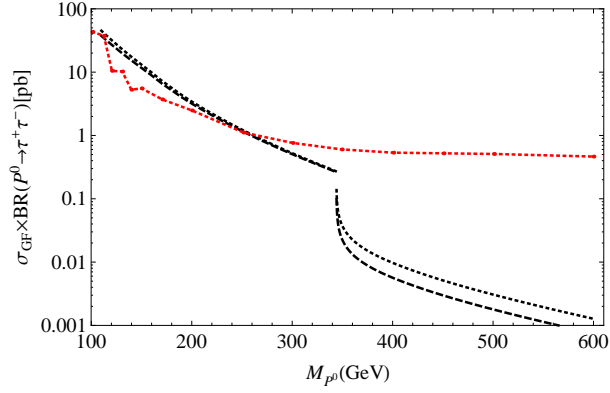}
\end{center}
\caption{ The cross section $\sigma_{\rm GF}( pp \to P^0)$ times branching ratio ${\rm BR}(P^0 \to \tau^+\tau^-)$ 
as a function of the $P^0$ mass in a high mass range up to 600 GeV for $N_{\rm TC}=3$ (dashed black) and 4 (dotted black) 
at the 7 TeV LHC in unit of pb. 
The red dotted line stands for the current 95\% CL upper bound from the ATLAS experiments with 1.06 
fb$^{-1}$~\cite{ATLAS-CONF-2011-132}. }  
\label{P0-tautau}
\end{figure}

On the other hand, the $P^0$ decays to fermion pairs get enhanced 
since the gluon fusion (GF) production cross section is highly enhanced (by about factor 10) 
due to techni-fermion loop contributions: 
\begin{equation} 
 \frac{\sigma(gg \to P^0)}{\sigma(gg \to h_{\rm SM})} 
 \approx 
 7 \left( \frac{N_{\rm TC}}{3} \right)^2 
 \,, \label{P0gg}
\end{equation}
where heavy quark mass limit for top and bottom quarks has been taken. 
Accordingly, the cross section times branching ratio for decays to the light fermion pairs are also enhanced  
compared to those of the SM Higgs. 
In fact, the current LHC data on the $\tau^+\tau^-$ channel severely constrains 
the $P^0$ mass to exclude it up to $m_{P^0} = 2 m_t$~\cite{Chivukula:2011ue}.  
 Note, however, that, as shown in the previous section, the typical value of $m_{P^0}$ estimated in the walking TC 
  exceeds the top pair threshold, i.e., $m_{P^0} = 397$ GeV (See Table~\ref{tab:TP}).  
 In such a higher mass region $m_{P^0} \ge 2 m_t$, the cross section $\sigma(pp \to P^0 \to \tau^+\tau^-)$ 
gets highly suppressed to be about order of $10^{-2}$ pb at around 397 GeV~\cite{Chivukula:2011ue}, 
which is well below the current 95\% C.L. upper bound $\sim 1$ pb at around 397 GeV~\cite{ATLAS-CONF-2011-132}. 
This is because the $t\bar{t}$ channel is open to be dominant. See Fig.~\ref{P0-tautau}.

In Ref.~\cite{Chivukula:2011ue} the analysis has been done with simple-minded Yukawa couplings for techni-pions assumed, 
namely, setting the overall factors associated with the pion currents, like ($1/2\sqrt{3}$, $3/2\sqrt{3}$) 
for the couplings  $(g_{P^0qq}, g_{P^0ll})$ in Eq.(\ref{P0-yukawa-int}), to unity. 
The present study has properly incorporated such factors specific to the one-family techni-pions, 
so that the current LHC limit from the $\tau^+\tau^-$ channel gets slightly modified as seen from Fig.~\ref{P0-tautau} 
to allow a small window below $2m_t$.

\begin{figure}

\begin{center} 
\includegraphics[scale=0.6]{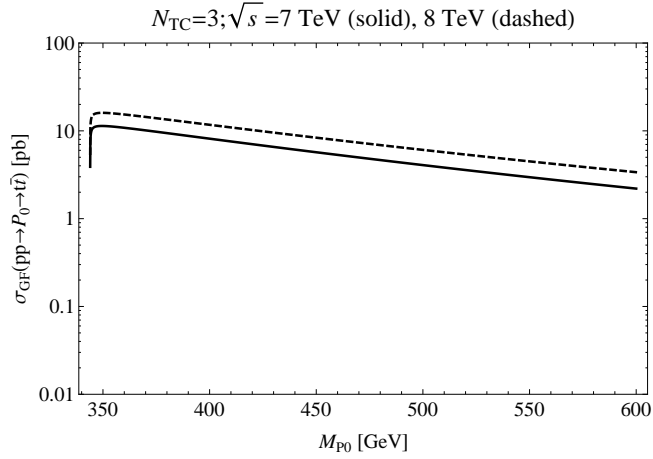}
\end{center}
\caption{ The $P^0$ contribution to the $t \bar{t}$ total cross section 
as a function of the $P^0$ mass for $N_{\rm TC}=3$ at $\sqrt{s}=7$ TeV (solid) and 8 TeV (dashed), 
in unit of pb. }  
\label{P0-ttbar-NWA}
\end{figure}

The branching fraction of $P^0$ with $m_{P^0}=397$ GeV is indeed 
governed by the $t \bar{t}$ mode, which is about 99\% 
compared to the SM Higgs case at around 397 GeV ${\rm BR}_{t \bar{t}}^{h_{\rm SM}} \simeq 15\%$~\cite{Dittmaier:2011ti}, 
but the total width remains as small as a few GeV, so it is still a narrow resonance.  
Thus the $P^0$ peak in the highly enhanced $t\bar{t}$ channel 
will be distinct to be measured by the $t \bar{t}$ invariant mass distribution~\cite{CMSPASTOP-11-009}.  
Figure~\ref{P0-ttbar-NWA} shows the $P^0$ contribution to the $t \bar{t}$ total cross section 
in the narrow width approximation as  a function of the $P^0$ mass in a range above $2 m_t$.   
 The $P^0$ resonance effect thus will yield about 8 pb to the $t \bar{t}$ total cross section 
 at the mass $m_{P^0}=397$ GeV, which is still within the current 1 sigma error of the $\sigma_{t\bar{t}}$ measurement 
at the LHC~\cite{Chatrchyan:2012vs} 
and hence will be tested more clearly by the upcoming 2012 data.

 \subsection{Isosinglet-color octet techni-pion $\theta_a$}

\begin{table}
\begin{tabular}{|c||cccccc|} 
\hline 
\hspace{10pt} $N_{\rm TC}$ \hspace{10pt} 
&\hspace{10pt} $m_{\theta_a}=449 \sqrt{\frac{3}{N_{\rm TC}}}$ [GeV] \hspace{10pt} 
&\hspace{10pt} $\Gamma_{\rm tot}^{\theta_a}$ [GeV]  \hspace{10pt} 
&\hspace{10pt} ${\rm BR}_{gg}$ \hspace{10pt} 
&\hspace{10pt} ${\rm BR}_{gZ}$ \hspace{10pt} 
&\hspace{10pt} ${\rm BR}_{g\gamma}$ \hspace{10pt} 
&\hspace{10pt} ${\rm BR}_{t \bar{t}}$ \hspace{10pt}  \\
 \hline \hline 
3 &  449 & 23 & $1.4 \times 10^{-2}$ & $3.1 \times 10^{-5}$  & $1.2 \times 10^{-4}$  & $9.8 \times 10^{-1}$  \\
4 &  389 & 14 & $2.5 \times 10^{-2}$ & $5.3 \times 10^{-5}$  & $2.2 \times 10^{-4}$  & $9.7 \times 10^{-1}$  \\
\hline 
&  \hspace{15pt} $r_{\rm GF}$   \hspace{15pt} 
&\hspace{15pt} $r_{\rm BR}^{gg}$ \hspace{15pt} 
&\hspace{15pt} $r_{\rm BR}^{t\bar{t}}$ \hspace{15pt}
&\hspace{15pt} $R_{t\bar{t}}$ \hspace{15pt} & & \\ 
 \hline \hline 
3 & 51 & 12 & 5.1 & 234 &  & \\ 
4 & 91 & 20 & 7.4 & 625  &  & \\
\hline 
\end{tabular}
\caption{
The $\theta_a$ total width, relevant branching fraction and numbers regarding the LHC signatures. 
Here we have defined: 
$r_{\rm GF} \equiv \sigma_{\rm GF}^{\theta_a}/\sigma_{\rm GF}^{h_{\rm SM}}$ with the gluon fusion production cross section at $\sqrt{s}=$ 7 TeV 
$\sigma_{\rm GF}$; $r_{\rm BR}^X \equiv {\rm BR}(\theta_a \to X)/{\rm BR}(h_{\rm SM} \to X)$; $R_X \equiv r_{\rm GF} \times r_{\rm BR}^X$. 
The branching ratios and 7 TeV LHC production cross section for the SM Higgs are taken from Ref.~\cite{Dittmaier:2011ti}.     
}
\label{tab:P8-signals} 
\end{table}

 From Eqs.(\ref{P8-gauge-int}) and (\ref{theta-yukawa-int}), 
the $\theta_a$ decay widths to the SM gauge boson and fermion pairs are calculated 
 to be  
\begin{eqnarray} 
\Gamma (\theta_a \to gg) 
&=&
\frac{5 N_{\rm TC}^2 \alpha_s^2 G_F m^3_{\theta_a}}{48 \sqrt{2} \pi^3} 
\,, \nonumber \\ 
\Gamma (\theta_a \to Zg) 
&=&
\frac{ N_{\rm TC}^2 \alpha_{\rm EM} \alpha_s  G_F m^3_{\theta_a} s^2}{72 \sqrt{2} \pi^3 c^2} 
\left( 1 - \frac{m_Z^2}{m_{\theta_a}^2} \right)^{3} 
\,, \nonumber \\ 
\Gamma (\theta_a \to \gamma g) 
&=&
\frac{ N_{\rm TC}^2 \alpha_{\rm EM} \alpha_s  G_F m^3_{\theta_a} }{72 \sqrt{2} \pi^3} 
\,, \nonumber \\ 
\Gamma (\theta_a \to q\bar{q}) 
&=&
\frac{G_F m_{\theta_a} m_q^2}{\sqrt{2} \pi} \left( 1 - \frac{4 m_q^2}{m_{\theta_a}^2} \right)^{1/2} 
\,. 
\end{eqnarray} 
The total width and branching ratios at $m_{\theta_a}=449 \, {\rm GeV}\, \sqrt{3/N_{\rm TC}}$ 
are shown in Table~\ref{tab:P8-signals} for $N_{\rm TC}=3, 4$.  
The GF dominates in 
the $\theta_a$ production process at the LHC, yielding the production 
cross section  enhanced by the QCD color factor $(N_c^2-1)$ 
and techni-fermion loop contributions, to be larger than that of the SM Higgs by about factor $10^2$: 
\begin{equation} 
 \frac{\sigma(gg \to \theta_a)}{\sigma(gg \to h_{\rm SM})} 
 = (N_c^2-1) \times \frac{\Gamma(\theta_a \to gg)}{\Gamma(h_{\rm SM} \to gg)}
 \approx 
 8 (N_c^2-1) \left( \frac{N_{\rm TC}}{3} \right)^2 
 \,, \label{P8gg}
\end{equation}
where $N_c=3$ and heavy quark mass limit for top and bottom quarks has been taken. 
The $\theta_a$ almost completely decays to $t\bar{t}$ pair with the partial decay rate 
$\Gamma(\theta_a \to t\bar{t}) (\simeq \Gamma_{\theta_a}^{\rm tot})|_{449\,{\rm GeV}} \simeq 23(14)$ GeV for $N_{\rm TC}=3(4)$, 
which is comparable  to that of the SM Higgs at the same mass. 
In spite of the large decay rate to $t \bar{t}$,  
the total width is small enough to treat the $\theta_a$ to be a narrow resonance.   
The $\theta_a$ is thus expected to give a large and sharp resonant contribution 
to the LHC $t\bar{t}$ events.

The $\theta_a$ contribution to the $t\bar{t}$ total cross section at $\sqrt{s}=7$ TeV 
in the narrow width approximation 
is estimated at the mass $m_{\theta_a} = 449\, {\rm GeV}$ for $N_{\rm TC}=3$ to be  
\begin{eqnarray} 
  \sigma_{t\bar{t}}^{\theta_a} \Bigg|_{m_{\theta_a}=449 {\rm GeV}}
  &\approx& 
  \sigma_{\rm GF}(pp \to \theta_a) \times {\rm BR}(\theta_a \to t\bar{t}) \Bigg|_{m_{\theta_a}=449 {\rm GeV}}
  \nonumber \\ 
  &\simeq& 60 \,{\rm pb} 
  \,, 
\end{eqnarray}
which is somewhat too large,  yielding 30\% of the presently observed cross section $\sigma_{\rm t\bar{t}} \simeq 180$ pb 
with the accuracy about 13\%~\cite{Chatrchyan:2012vs}.     
The current LHC data on the $t\bar{t}$ cross section thus require the 
$\theta_a$ mass to be below the threshold for the top quark pair, namely, $N_{\rm TC} \ge 6$.

\chushi{
 \begin{figure}

\begin{center} 
\includegraphics[scale=0.6]{P8-ttbar-NWAtotcross.eps}
\end{center}
\caption{ The $\theta_a$ contribution to the $t \bar{t}$ total cross section 
as a function of the $\theta_a$ mass for $N_{\rm TC}=3$ (dashed black) and 4 (dotted black), 
in unit of pb. }  
\label{P8-ttbar-NWA}
\end{figure}
}

\begin{figure}

\begin{center} 
\includegraphics[width=8.5cm]{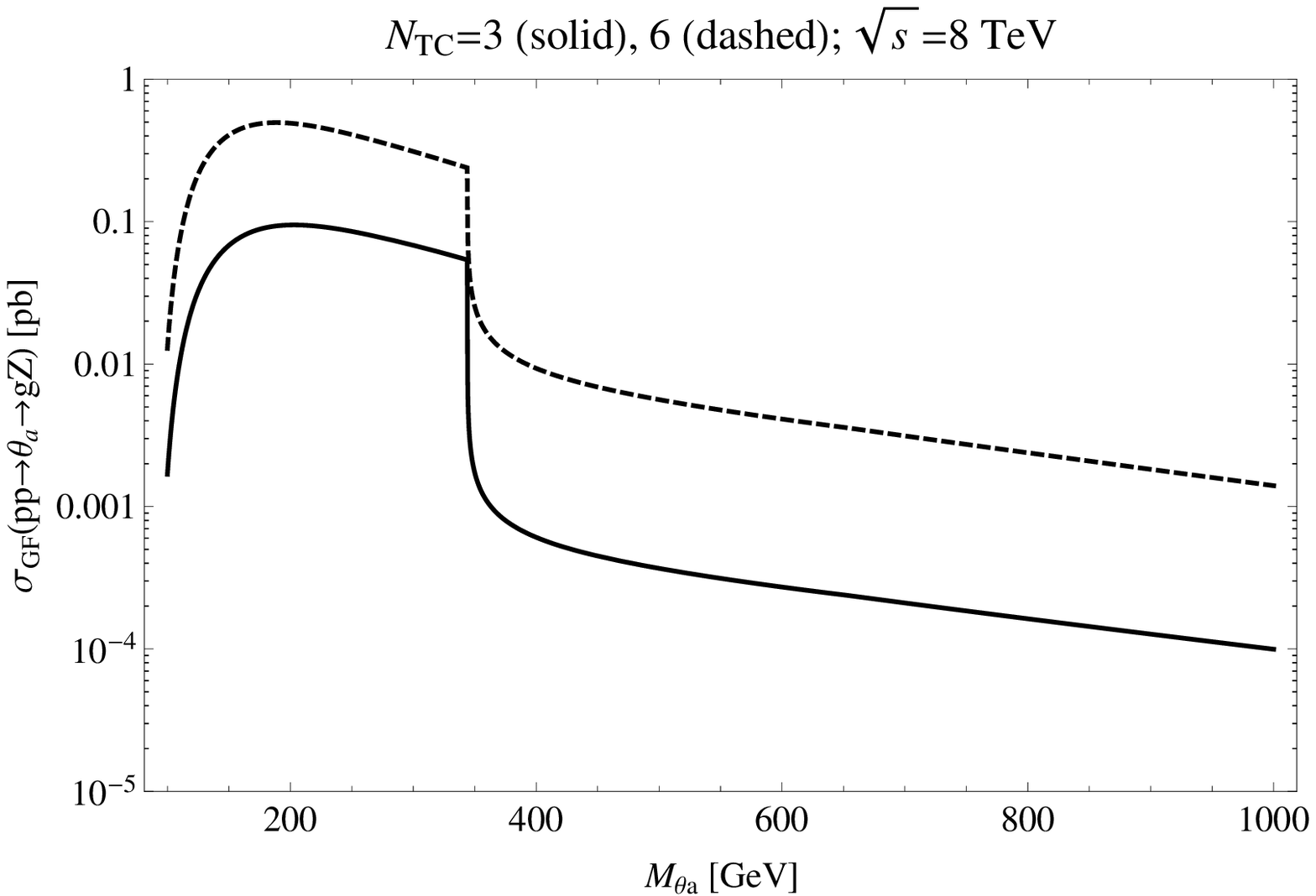}
\includegraphics[width=8.5cm]{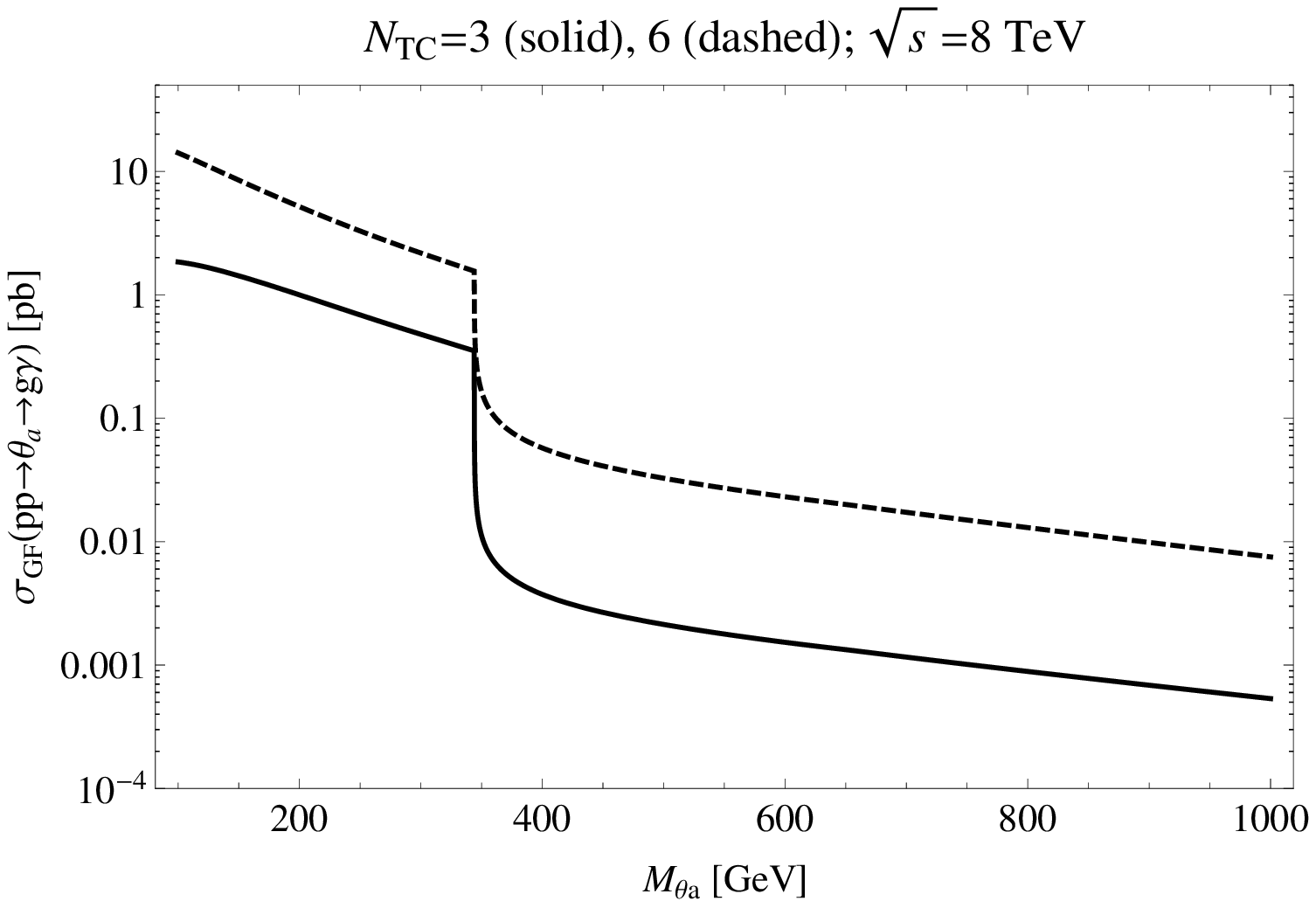}
\end{center}
\caption{ The predicted $\theta_a$ contribution to the LHC cross sections $\sigma(pp \to gZ)$ (left panel) and 
$\sigma(pp \to g\gamma)$ (right panel)
as a function of the $\theta_a$ mass for $N_{\rm TC}=3$ (solid) and 6 (dashed) with $\sqrt{s} = 8$ TeV fixed,  
in unit of pb. }  
\label{P8-gZ-NWA}
\end{figure}

Another interesting discovery channel for the $\theta_a$ would be 
the $\theta_a \to gZ/\gamma$ mode~\cite{Chivukula:1995dt}. 
The cross section may be evaluated by assuming the GF dominance and 
taking the narrow width approximation:  
\begin{eqnarray} 
\sigma_{\rm GF}(pp \to \theta_a \to gZ/\gamma) 
&\approx &
\sigma_{\rm GF}(pp \to \theta_a) \times {\rm BR}(\theta_a \to gZ/\gamma) 
\nonumber \\  
&=& 
\frac{32 \pi^2}{s}
\int d \eta \, 
 f_{g/P}(\sqrt{\tau} e^\eta, m_{\theta_a}^2) f_{g/P}(\sqrt{\tau} e^{-\eta}, m_{\theta_a}^2) 
\cdot {\cal C}_{Z/\gamma} \cdot 
\frac{\Gamma(\theta_a \to gg) \, {\rm BR}(\theta_a \to gZ/\gamma)}{m_{\theta_a}} 
\,, \label{gZ:cross}
\end{eqnarray}
with a pseudorapidity $\eta$ cut fiducially at $|\eta|=2.5$~\cite{Aad:2009wy} and 
the parton distribution function $f_{g/P}$ from CTEQ6~\cite{Pumplin:2002vw}. 
Here ${\cal C}_{Z/\gamma}$ denotes the multiplication factor for the initial, resonance and final states 
for the partonic cross section, i.e., 
${\cal C}_{Z/\gamma} = (\frac{1}{8} \cdot \frac{1}{8})_{gg} \times (8)_{\theta_a} \times (\frac{1}{8} \cdot \frac{1}{3}(\frac{1}{2}))_{gZ/\gamma}$. 
In Fig.~\ref{P8-gZ-NWA} the predicted cross section is plotted as a function of the 
$\theta_a$ mass $m_{\theta_a}$. 
 For the reference value $m_{\theta_a}=449 \, {\rm GeV} \sqrt{3/N_{\rm TC}}$ with $N_{\rm TC}=6$,  
the expected total number of the $gZ(\gamma)$ events will roughly be a few (ten) thousands for 5 fb$^{-1}$ data at 
$\sqrt{s} = 8$ TeV.  
The estimate of SM background and significance for the discovery will be pursued in another publication.

\section{Summary}  
\label{summary}

We have explicitly computed the techni-pion masses in the Farhi-Susskind one-family model 
taking into account essential features of walking TC.  
The explicit estimate of the masses was done by using recent results on 
a nonperturbative analysis based on 
the ladder Schwinger-Dyson equation employed in a modern version of walking TC.

The charged pion masses 
were calculated by evaluating one-EW gauge boson exchange diagrams, 
to show that 
the collected contributions take the form of integral over the momentum square $Q^2$ 
with respect to difference between vector and axialvector current correlators $\Pi_{V-A}$, 
similarly to computation for charged pion mass in QCD. 
The EW gauge boson contributions were shown to dramatically cancel   
each other, so that there are no sizable corrections to the masses, 
although the $\Pi_{V-A}$ is quite sensitive to the walking dynamics.

In contrast, sizable corrections were seen in the one-gluon exchange diagram 
yielding the colored techni-pion masses. 
We found that the size of correction is actually enhanced by a large logarithmic factor 
$\log \Lambda_{\rm TC}/F_\pi$, compared to the naive-scale up version of TC. 
This is due to the characteristic ultraviolet scaling of $\Pi_{V-A}$ 
in the walking TC, which can be seen in the asymptotic form of $\Pi_{V-A}$ 
for ultraviolet region through the slow damping behavior, $\Pi_{V-A} \sim 1/Q^{4-2\gamma_m}$.

We also evaluated an ETC-induced four-fermion interaction breaking separate 
chiral symmetry between techni-quarks and -leptons, which gives the masses to techni-pions 
coupled to the separate chiral currents. 
The masses were shown to be enhanced due to the chiral condensate enhanced by 
the large anomalous dimension.

It then turned out that all the techni-pions are on the order of several hundred  GeV (See Table~\ref{tab:TP}).

Based on our estimation, we finally discussed the phenomenological implications to the LHC signatures, 
focusing on neutral isosinglet techni-pions ($P^0$ and $\theta_a$), in comparison with the SM Higgs.  
We found the characteristic LHC signatures can be seen through excessive top quark productions 
for both two techni-pions. 
More on the techni-pion LHC studies is to be pursued in the future.

\section*{Acknowledgments}

This work was supported by 
the JSPS Grant-in-Aid for Scientific Research (S) \#22224003, (C) \#23540300 (K.Y.) 
and \#23-01781 (J.J.). 
J. J. is also supported by the JSPS Postdoctoral Fellowships for Foreign Researchers P11781.

\appendix 
\renewcommand\theequation{\Alph{section}.\arabic{equation}}

\section{Techni-dilaton $\phi$}  
\label{sec:TD}

In this Appendix we shall briefly address the phenomenological contributions to techni-dilaton   
signatures coming from the techni-pion couplings.

 From Eq.(\ref{Lag:int}) we read off the techni-dilaton couplings to the SM particles and techni-pions. 
 The formulas for decays to the SM particles were previously reported in Ref.~\cite{Matsuzaki:2011ie,Matsuzaki:2012gd,Matsuzaki:2012fq}.   
 While the partial width for two-body decay to techni-pions is calculated as 
\begin{equation} 
 \Gamma(\phi \to P^A P^B) 
 = \delta^{AB} \frac{M_{\phi}^3}{32 \pi F_{\phi}^2} \left( 1 - \frac{4 m_{P^A}^2}{M_{\phi}^2} \right)^{1/2}
\,, \label{width:TD-TP}
\end{equation}
 where $A,B$ denote labels of techni-pions used as in Table~\ref{tab:TP} 
and we have set $\gamma_m \simeq 1$. 
In deriving Eq.(\ref{width:TD-TP}) we 
added the appropriate techni-pion mass terms in Eq.(\ref{Lag:int}).

\begin{figure}
\begin{center}
   \includegraphics[scale=0.5]{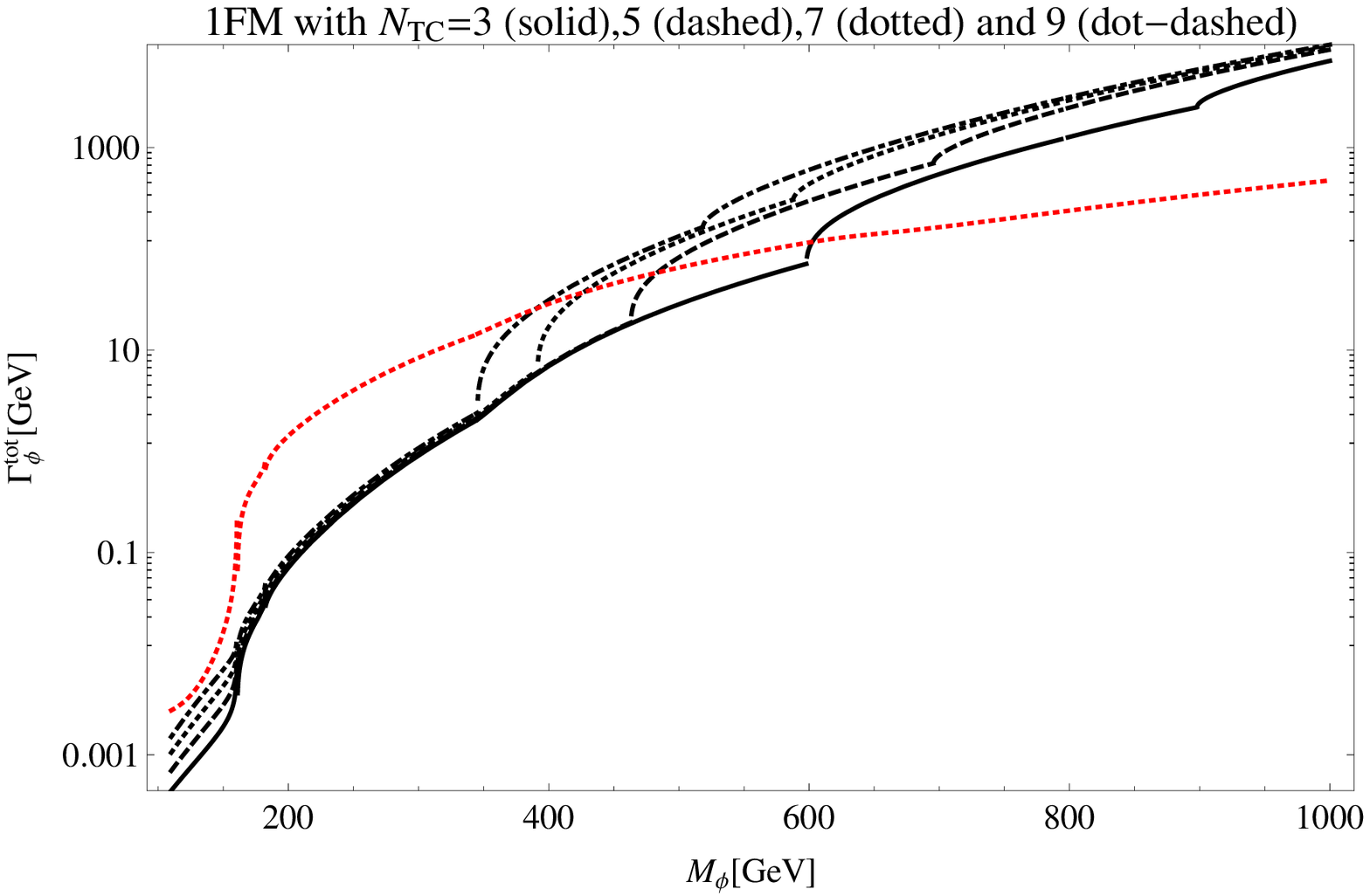}
\caption{ The total width of techni-dilaton as a function of $M_{\phi}$ in unit of GeV 
drawn by black solid ($N_{\rm TC}=3$), dashed ($N_{\rm TC}=4$), dotted ($N_{\rm TC}=5$) and dot-dashed ($N_{\rm TC}=6$) curves,  
in comparison with that of the SM Higgs (red dotted curve). 
Use has been made of 
the reference values of techni-pion masses listed in Table~\ref{tab:TP} for $\Lambda_{\rm TC}=10^3$ TeV.  
\label{TD-tot-width-full}
}
\end{center} 
 \end{figure}

 Combining Eq.(\ref{width:TD-TP}) with the previously reported ones for the SM particles 
 and using $m_F = 319 \, {\rm GeV} \sqrt{\frac{3}{N_{\rm TC}}}$ and 
$F_{\phi}=383 \, {\rm GeV} (\frac{600\,{\rm GeV}}{M_\phi})$~\cite{Matsuzaki:2011ie,Matsuzaki:2012gd,Matsuzaki:2012fq}, 
in Fig.~\ref{TD-tot-width-full} we plot the total width as a function of $M_{\phi}$ with $N_{\rm TC}=3$ taken, 
in comparison with the SM Higgs case.  Also have been used  
the reference values of techni-pion masses listed in Table~\ref{tab:TP}.  
 Looking at this figure, we see that the total width becomes much larger than that of the SM Higgs 
at around 500 GeV. 
This happens because the decay channel for color-triplet techni-pion pair starts to be kinematically allowed.

\begin{figure}
\begin{center}
   \includegraphics[scale=0.6]{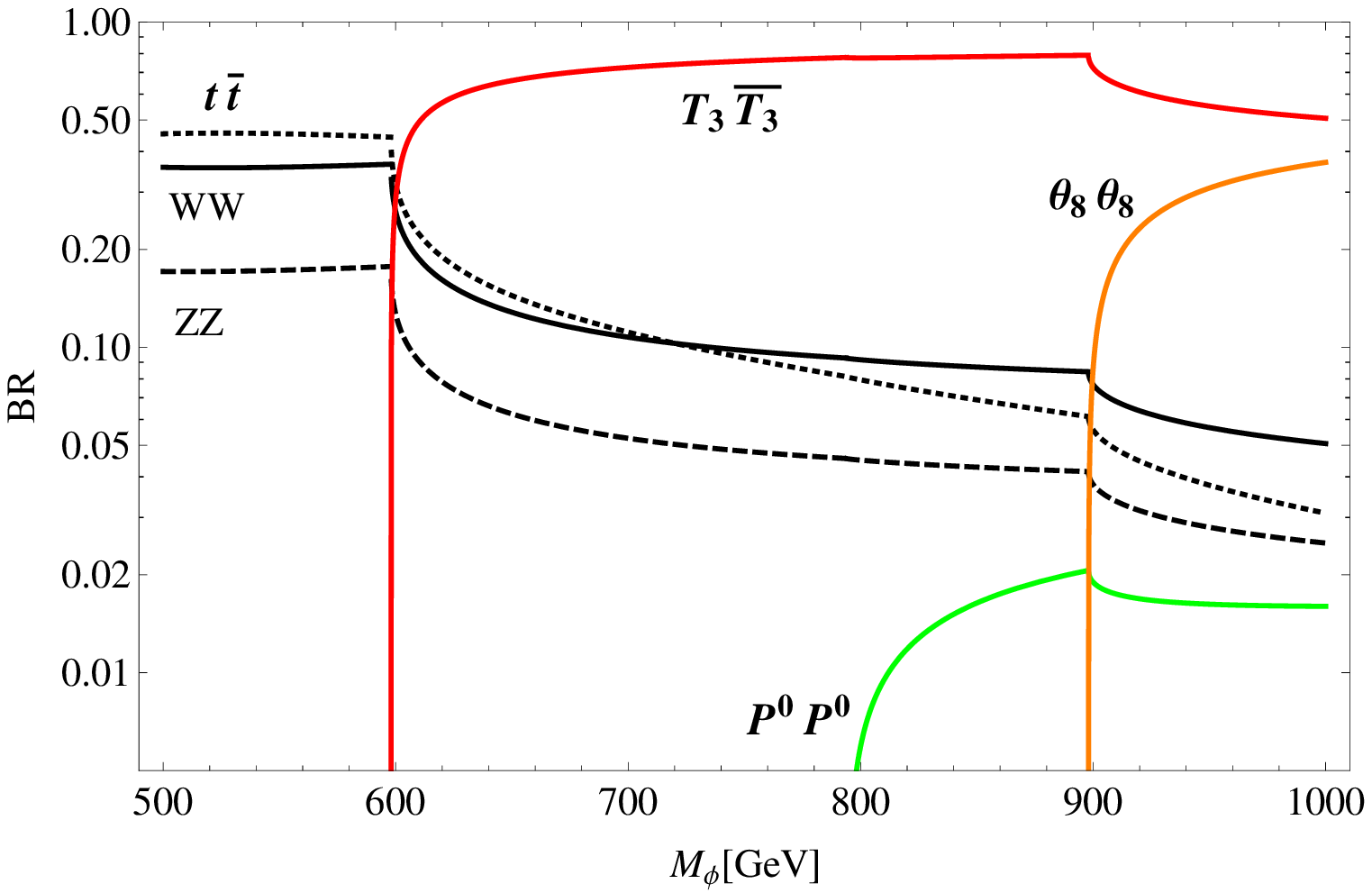}
\caption{ The techni-dilaton branching fraction for the higher mass range $500 \le M_{\phi} \le 1000$ GeV. 
The number of TC $N_{\rm TC}$ and $\Lambda_{\rm TC}$ have been taken to be 3 and $10^3$ TeV, respectively. 
 The color-triplet $(T_c^i, T_c)$ and -octet ($\theta_a^i,\theta_a$) techni-pions are collectively expressed as $T_3$ and $\theta_8$, 
respectively. 
The techni-pion masses are set to the reference values listed in Table~\ref{tab:TP}. 
\label{TD-BR}
}
\end{center} 
 \end{figure}

The branching fraction in fact becomes dramatically changed above around 500 GeV 
since a new decay channel to the lightest techni-pion $T_c\bar{T}_c$ pair (See Table~\ref{tab:TP}) starts to 
open to be dominant.  
In Fig.~\ref{TD-BR} we show the branching fraction for mass range $500 \le M_{\phi} \le 1000$ GeV 
taking $N_{\rm TC} =3$ and $\Lambda_{\rm TC}=10^3$ TeV. 
For the associated LHC signatures of techni-dilaton, see Refs.~\cite{Matsuzaki:2011ie,Matsuzaki:2012gd,Matsuzaki:2012fq}.

\end{document}